%% file: paper.tex
\documentclass{vgtc}                          % final (conference style)
\ifpdf%                                % if we use pdflatex
  \pdfoutput=1\relax                   % create PDFs from pdfLaTeX
  \usepackage{graphicx}                % allow us to embed graphics files
  \DeclareGraphicsExtensions{.pdf,.png,.jpg,.jpeg} % for pdflatex we expect .pdf, .png, or .jpg files
\else%                                 % else we use pure latex
  \usepackage{graphicx}                % allow us to embed graphics files
  \DeclareGraphicsExtensions{.eps}     % for pure latex we expect eps files
\fi%

\graphicspath{{figures/}{pictures/}{images/}{./}} % where to search for the images

\usepackage{microtype}                 % use micro-typography (slightly more compact, better to read)
\PassOptionsToPackage{warn}{textcomp}  % to address font issues with \textrightarrow
\usepackage{textcomp}                  % use better special symbols
\usepackage{mathptmx}                  % use matching math font
\usepackage{times}                     % we use Times as the main font
\usepackage{cite}                      % needed to automatically sort the references
\usepackage{tabu}                      % only used for the table example
\usepackage{booktabs}                  % only used for the table example
\usepackage[super]{nth}
\usepackage[]{algorithm2e}
\usepackage{gensymb}

\usepackage{xcolor}
\usepackage{lipsum}
\usepackage[colorinlistoftodos]{todonotes}
\usepackage{float}

\usepackage{comment}
\usepackage{amsmath}
\usepackage[switch]{lineno}

\definecolor{myblue}{rgb}{0.0156, 0.2431, 0.5843}
\definecolor{gray}{rgb}{0.7,0.7,0.7}
\definecolor{orange}{rgb}{1.0,0.5,0.0}

\newcommand*{\rv}{\textcolor{black}}

\newcommand{\summary}[1]{{\textcolor{orange}{#1}}} 
\renewcommand{\summary}[1]{}

\onlineid{3720}

\vgtccategory{Research}

\vgtcinsertpkg

\title{Context-Responsive Labeling in Augmented Reality}

\author{
Thomas K{\"o}ppel \thanks{e-mail: tkoeppel@cg.tuwien.ac.at}\\ %
\scriptsize TU Wien, Austria %
\and 
M. Eduard Gr{\"o}ller\thanks{e-mail: groeller@cg.tuwien.ac.at}\\ %
\scriptsize TU Wien, Austria \\
\scriptsize VRVis Research Center, Austria %
\and 
Hsiang-Yun~Wu\thanks{e-mail: hsiang.yun.wu@acm.org}\\ %
\scriptsize TU Wien, Austria %
}

\teaser{
\centering{
 \setlength{\tabcolsep}{0.5pt}
 \begin{tabular}{cccccc}
    \includegraphics[width=0.13\linewidth]{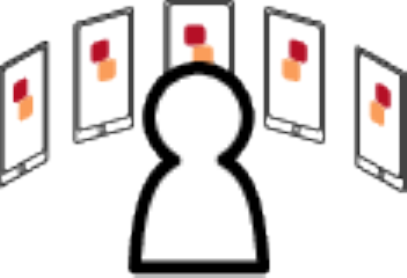} &
    \includegraphics[width=0.17\linewidth]{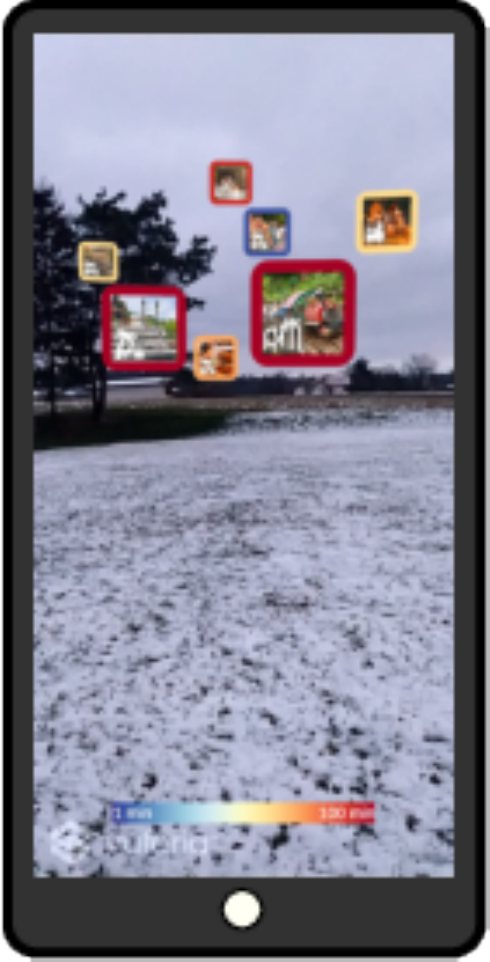} &
    \includegraphics[width=0.17\linewidth]{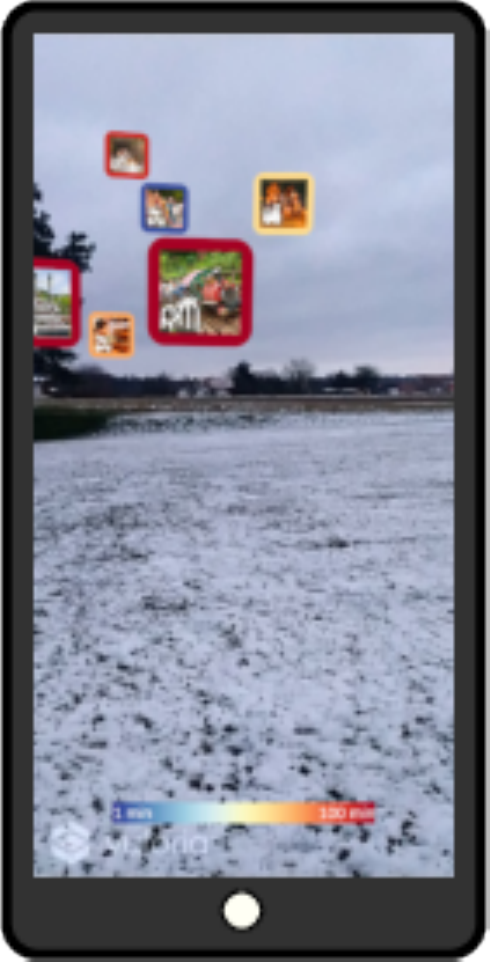} &
    \includegraphics[width=0.17\linewidth]{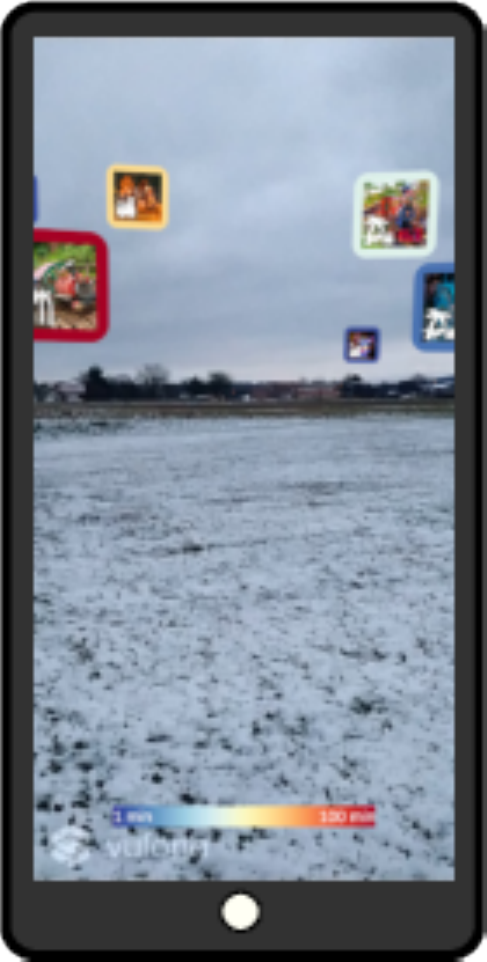} &
    \includegraphics[width=0.17\linewidth]{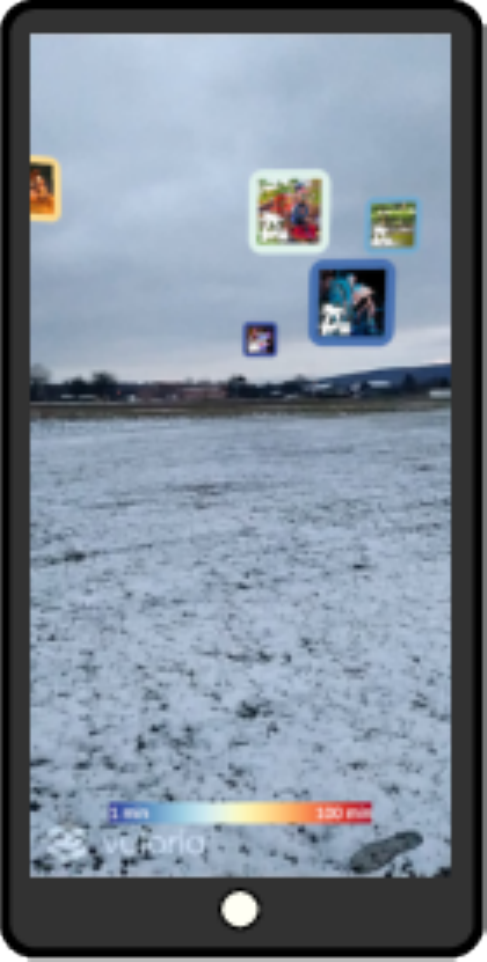} &
    \includegraphics[width=0.17\linewidth]{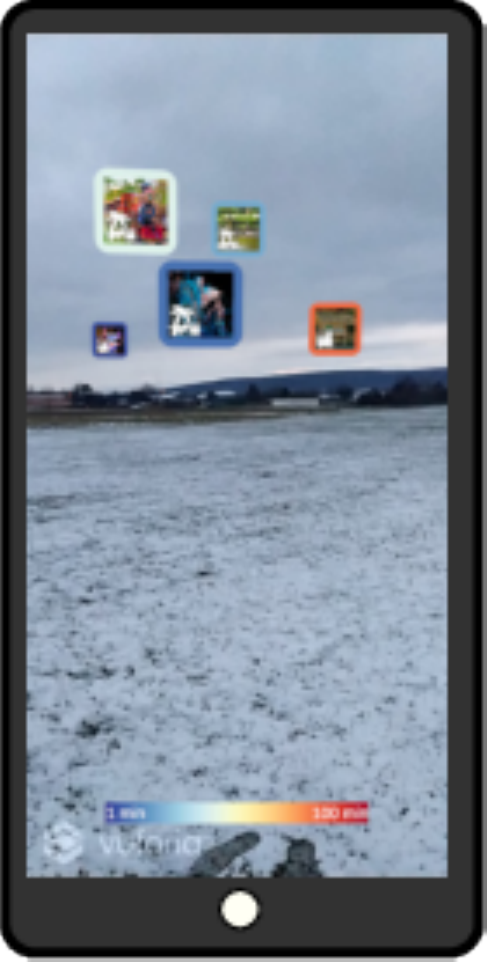}  \\
   & \rv{(a) $-60^{\circ}$}&\rv{(b) $-30^{\circ}$}&\rv{(c) $0^{\circ}$}&\rv{(d) $30^{\circ}$}&\rv{(e) $60^{\circ}$}\\ 
 \end{tabular}
}
\caption{Responsive labeling of the \emph{Tokyo Disneyland Dataset} in AR. The images present results of different viewing angles to show consistent label positions when the AR device is rotated toward different directions by the user.}
\label{fig:teaser}
}

\input{sections/abstract}

\begin{document}
% need to be removed
% \linenumbers


\maketitle

\input{sections/intro}

\input{sections/related}
\input{sections/overview}
\input{sections/arsetting}
\input{sections/occlusion}

\input{sections/lod}

\input{sections/transition}
\input{sections/result}

\input{sections/evaluate}
\input{sections/limitations}
\input{sections/conclude}

\acknowledgments{
Part of the research was enabled by VRVis funded in COMET (879730) a program managed by FFG.
}

\bibliographystyle{abbrv-doi}

\bibliography{paper}
\end{document}

%% file: sections/abstract.tex
\abstract{
\rv{Route planning and navigation are common tasks that often require} additional information on points of interest.
Augmented Reality (AR) enables mobile users to \rv{utilize} text labels, in order to provide a composite view associated with additional information in a real-world environment. 
\rv{Nonetheless,} displaying all labels for points of interest on a mobile device will lead to unwanted overlaps between information, and thus a context-responsive strategy to properly arrange labels is expected.
\rv{The technique should remove overlaps, show the right level-of-detail, and maintain} label coherence. 
This is necessary as the viewing angle in an AR system may change rapidly due to users' behaviors. \rv{Coherence} plays an essential role in retaining user experience and knowledge, as well as avoiding motion sickness. 
In this paper, we develop an approach that systematically manages label visibility and levels-of-detail, \rv{as well as eliminates} unexpected incoherent movement. 
We introduce three label management strategies, including (1) \emph{occlusion management}, (2) \emph{level-of-detail management}, and (3) \emph{coherence management} by balancing the usage of the mobile phone screen. 
A greedy approach is developed for fast occlusion handling \rv{in AR}. A level-of-detail scheme is adopted to arrange \rv{various types of labels}. A 3D scene manipulation \rv{is then} built to simultaneously suppress the incoherent behaviors induced by viewing angle changes. 
Finally, we present the feasibility and applicability of our approach \rv{through} one synthetic and two real-world scenarios, followed by a qualitative user study.
} % end of abstract

\CCScatlist{
  \CCScatTwelve{Human-centered computing}{Visu\-al\-iza\-tion}
}

%% file: sections/intro.tex
\section{Introduction}
\label{sec:intro}

% high-level background
We schedule and \rv{plan routes} irregularly in our everyday life. 
For example, we visit offices, go to restaurants, or see doctors, in order to accomplish necessary tasks. 
In some cases, such as visiting medical doctors or popular restaurants, one has to wait in a queue until being able to \rv{proceed}. 
\rv{This is time-inefficient and most people try to avoid it.}
% especially in a pandemic situation (e.g., COVID-19 pandemic).
Normally, if a person needs to decide the next place to visit, he or she can extract knowledge about the targets of interest. Then a decision is made based on the corresponding experience or referring locations using a map.
2D maps are one of the most popular methods that describe the geospatial information \rv{of objects, to give an overview of the object %relative 
positions in a certain area.
\rv{With} a 2D map for navigation, users need to remap or translate the objects on the map to the real environment, to} understand the relationships and distances to these objects~\cite{guarese}.
This inevitably \rv{strains} our cognition. It is also the reason why some people cannot quickly \rv{locate} themselves on \rv{a} 2D map or find the correct direction or orientation immediately.
\rv{\emph{Augmented Reality (AR)} and \emph{Mixed Reality (MR)} have} been proposed to overlay information directly \rv{on} the real-world environment with a lower complexity by instructing users \rv{in an effective way% the most effective condition
~\cite{McMahon:2015:JSET, Ens:2019:JHCS}. 
In this paper, we use AR \rv{as our technique of choice} for the explanation.
Displaying texts or images in \rv{AR or MR} allows us to acquire information encoded with geotagged data and stored in GISs.
It is also known that using AR for guiding users in exploring the real environment can be more effective in comparison to a 2D representation~\cite{Devaux:2018:IV}.}
% Based on these observations, we hypothesize that \rv{using labels in AR} for guiding users in exploring the real environment can be more effective in comparison to a 2D representation. 

\rv{In mixed environments}, \emph{points of interest (POIs)} are often associated with text labels~\cite{hedgehog, imagebased, nextgen} in order to present additional information (e.g., name, category, etc.). For example, an Augmented Reality Browser (ARB) facilitates us to embed and show relevant data in a real-world environment. 
% To have better association with POIs in AR, real-world objects are usually annotated with text labels  showing the name or the view of a POI. 
Technically, POIs are registered at certain geographical positions via GPS coordinates. 
Based on the current position and the viewing angle of the device, the POIs are annotated and the corresponding labels are then projected to the screen of the user's device. Naive labeling strategies \rv{can} lead to occlusion problems between objects, especially in an environment with a dense arrangement of POIs. 
Additionally, properly selecting the right level of a label to present information can help to avoid overcrowded \rv{situations}.
Moreover, retaining the consistency between \rv{successive} frames also enables us to maintain a good user experience and to avoid motion sickness.
% Furthermore, occlusions in AR are often solved in 2D \cite{imagebased, grassetimage, labelsurvey}, which may lead to another problem, as 3D temporal behavior is not obvious in 2D space as concluded by Tatzgern~et~al.~\cite{hedgehog}. 
Based on the aforementioned findings, we summarize that a good AR labeling framework should address: 
\begin{itemize}
    \item[\textbf{(P1)}] \textbf{The occlusion of densely placed labels in AR space.} Occlusion removal has been considered as a primary design criterion \rv{in visualization approaches. It reflects user preferences and also allows the system} to present information explicitly~\cite{Wu:2013:EuroVis}.
    \item[\textbf{(P2)}] \textbf{\rv{Limited} Information provided by plain text.} As summarized by Langlotz~et~al.~\cite{nextgen}, labels in AR often \rv{contain} plain text rather than other richer content, such as figures or hybrids of texts and figures.
    \item[\textbf{(P3)}] \textbf{Label incoherence due to the \rv{movement} of mobile devices.} During the interaction with \rv{an AR system}, the user may frequently change positions or viewing angles. This leads to unwanted flickering that impacts \rv{information consistency}~\cite{imagebased}.
\end{itemize}

% As stated by Langlotz~et~al~\cite{nextgen}, labels usually present textual information rather than "rich content", which represents our second problem: \textit{P2 Information provided by plain text labels is limited}. In addition, dense labeling leads to visual clutter \cite{nextgen} and to largely occupied screenspace \textbf{P3 Solving of P1 and P2 requires space which is limited on a mobile device (Section~\ref{subsec:module2})}. Due to the frequent position and viewing angle changes of the user during the interaction with the environment in AR, jittering represents another problem that influences the coherent information gathering of users \cite{imagebased}. When using 2D print outs for navigation, users need to map the objects they are looking at on the map to the real world  \cite{arvs2d}, which costs time and effort - \textit{P4 Navigation can be difficult (Section~\ref{subsec:module1})}.
%translate the map to the real world

In this paper, we develop a context-responsive framework to optimize label placement in AR. 
\rv{By \emph{context-responsive}, we refer to taking contextual attributes, such as GPS positions, mobile orientations, etc., into account. The system responds to the user with an appropriate positioning of labels.
The} approach contains three major components: (1) \emph{occlusion management}, (2) \emph{level-of-detail management}, and (3) \emph{coherence management}, which are essential for the approach to be context-responsive. 
%since we aim to balance these three components by optimizing the screen space usage of mobile devices~\cite{Hoffswell:2020:CHI}.
\rv{The} \emph{occlusion management} eliminates overlapping labels by adjusting the positions of occluded labels with a greedy approach to achieve a fast performance.
Then, a levels-of-detail scheme is introduced to select the \rv{appropriate} level in \rv{a} hierarchy and present it based on how densely \rv{packed} the labels are in the view volume of the user. 
We construct a 3D scene to manipulate and control the movement of labels \rv{enhancing} the user experience.
%

% Contribution
We introduce a novel approach to manage label placement tailored to AR. It enables an interactive environment \rv{with} continuous changes of device positions and orientations. 
A survey by Preim~et~al.~\cite{preim1} concluded that existing labeling techniques often \rv{resolve} overlapping labels once the camera stops moving or the camera position is assumed to be fixed to begin with. 
\rv{Approaches} often project labels to a 2D plane to \rv{determine} the occlusions and \rv{then} perform occlusion removal.
% Therefore, these techniques are not suitable to use in AR where the device is always moving. 
However, object movement in 3D is not obvious in the 2D projections of a 3D scene, which leads to temporal inconsistencies that \rv{are} harmful to label readability~\cite{hedgehog}. 
\rv{{\v C}mol{\'i}k et al.~\cite{Cmolik:TVCG:2020} summarized the difficulty of retaining label coherence due to many discontinuities of objects projected into 2D images.}
As \rv{in the} sequence of snapshots \rv{in} Figure~\ref{fig:teaser},
we treat labels as objects in a 3D scene and \rv{apply} our management strategies \rv{for better quality control.}
%
%\todo{Need to revisit.}
%
% We solve occlusions of these labels by shifting the labels along the vertical axis as we consider the $x$ and $z$ coordinates crucial for navigational purposes. The $x$ and $z$ axes encode the latitude and longitude coordinates while the $y$ axis represents the height. We assume that close labels are more important than distant ones. According to the current position of the user, close labels are shifted less during the occlusion handling than distant ones as we treat labels as objects in the 3D scene.
%
In summary, \rv{the} main technical contributions are:
\begin{itemize}
    \item A fast label occlusion removal technique for mobile devices. 
    \item A clutter-aware level-of-detail management. 
    \item A 3D object arrangement that retains label coherence.
    \item \rv{A prototype to demonstrate the applicability of our approach~\cite{Koeppel:2021:repo}}.
\end{itemize}

% In order to overcome incoherent behavior due to the changes of viewing angles, we decide to resolve occlusions in AR in 3D. Our occlusion handling approach is computed based on the current position of the user and solves occlusions for each label around the user. This way, even abrupt viewing angle changes do not influence our occlusion-free result because these changes are considered during the calculation of our occlusion-free result (see Section~\ref{chap::method}). 

The remainder of \rv{the} paper is structured as follows: 
Section~\ref{sec:related} presents previous work and relates our approach \rv{to} existing research. 
An overview of our design principles and system \rv{is} described in Section~\ref{sec:overview}. In Section~\ref{sec:method}, we detail the methodology and technical \rv{aspects}. The implementation is explained and use cases are demonstrated in Section~\ref{sec:result}, followed by \rv{an evaluation} in Section~\ref{sec:evaluate}. The limitations are explained in Section~\ref{ssec:limitations}, and we conclude this work and provide future research directions in Section~\ref{sec:conclude}.
% We show the functionality of our tool, by incorporating a Tokyo Disneyland dataset as an example (Section~\ref{chap:scenario}) and present results for our scenario in Section~\ref{sec:result}. Section~\ref{chap::impldetails} highlights implementation details while Section~\ref{sec:evaluate} explains our evaluation. At the end of the paper, we describe the limitations (Section~\ref{chap::limitaitons}), and we conclude our paper and explain possible future improvements (Section~\ref{chap::conclusion}).

%% file: sections/related.tex
\section{Related Work}
\label{sec:related}

\rv{We} present a novel responsive approach \rv{considering} label occlusion, visual clutter, and coherence simultaneously. We \rv{discuss} related work to identify our contributions \rv{by first covering general navigation techniques, and then specific labeling topics in different applications and spaces.}

\subsection{Spatial Identification and Navigation}
\label{ssec:navi}

Spatial cognition studies show how people acquire experience and knowledge to identify where they are, how to continue the journey, and visit places effectively~\cite{Waller:2012:APA}.
\rv{Maps are classical tools used to detect positions and extract spatial information throughout human history~\cite{wu:2020:eurovis}, 
while modern maps often use markers to identify and highlight the \rv{locations} of POIs.}
2D maps may not be always optimal since the 2D information needs to be translated to the real environment~\cite{guarese}.

An alternative, or maybe a more intuitive way, is to map the information \rv{directly to the physical environment.
McMahon et al.~\cite{McMahon:2015:JSET} compared paper maps and Google Maps to AR or more specifically hand-held AR~\cite{Sereno:2021:TVCG}, which better supports people in terms of activating their navigation skills.}
\rv{Willett et al.~\cite{Willett:2017:TVCG} introduced embedded data representations, a taxonomy describing the challenges of showing data in physical space, and mentioned that occlusion problems have not yet been fully resolved.
}
Bell et al.~\cite{firstnavigation} proposed \rv{a} pioneering view-management approach to project objects \rv{onto the} screen while resolving occlusions or to arrange similar objects close to each other.
Guarese and Maciel~\cite{guarese} investigated MR, to assist navigation tasks by overlaying the real environment with virtual holograms. 
Schneider et al.~\cite{schneider} investigated an AR navigation concept, where the system projects the content onto a vehicle’s windshield to assist driving behaviors.

\subsection{Labeling in Various Spaces (2D, 3D, VR, and AR)}
\label{ssec:labeling}

Labeling is an automatic approach to position text or image labels in order to efficiently communicate additional information about POIs. It improves clarity and understandability of the \rv{underlying information~\cite{labelsurvey}.}
\rv{Internal labels} are overlaid onto their reference objects. External labels are placed outside the objects and are connected to them by leader lines.
Recently, {\v C}mol{\'i}k~et~al.~\cite{Cmolik:TVCG:2020} have introduced \emph{Mixed Labeling} that facilitates the integration of internal and external labeling \rv{in 2D}.
Labeling techniques have been extensively investigated in geovisualization, where resolving occlusions
and leader crossings~\cite{Lin:2010:pvis} are primary aesthetic criteria to ensure good readability.
Besides 2D labeling, in digital map services, such as Google Maps \rv{and} other GISs, scales have been considered to improve user interaction.
Active range optimization, for example, uses rectangular pyramids to \rv{eliminate} label-placement conflicts across different zoom levels~\cite{Been:2010:cg,Wu:2017:EuroVis}.
Labeling of 3D scenes has been mainly investigated in medical applications~\cite{Oeltze:2014:vcbm}, usually focusing on complex mesh and volume scenes\rv{, as well as} intuitiveness for navigation.
\rv{
Maass and D{\"o}llner~\cite{Maass:2006:WSCG} developed a labeling technique to dynamically attach labels to the hulls of objects in a 3D scene. Later they extended this billboard concept by taking occlusion with labels and scene elements into account~\cite{Maass:2008:CAG}.
} The approach by Kou{\v r}il et al.~\cite{kouril-2018-LoL} \rv{annotates} a complex 3D scene, involving multi instances across multiple scales in a dense 3D biological environment.
\rv{Occlusion in these approaches is detected after projecting objects into 2D. It is hard to maintain coherence.}

\rv{Handheld} Augmented Reality has become useful as the computing power of mobile devices \rv{has increased}. 
One advantage of using AR is to overlay information directly \rv{on} the real world that the user is familiar with. 
\rv{
For example, White and Feiner~\cite{White:2009:CHI} proposed \emph{SiteLens}, a situated visualization that embeds relevant data of the POIs in AR.
Veas et al.~\cite{Veas:2012:TVCG} investigated outdoor AR applications, where they focused on multiple-view coordination and occlusion with objects in the background. 
Labels are not fully researched here. 
}
As referred to in most of the \rv{following} papers, occlusions between labels have been considered as a primary issue in AR applications~\cite{nextgen, grassetimage,imagebased}. 
Grasset~et~al.~\cite{grassetimage} proposed a view management technique to annotate landmarks \rv{in} an image. 
\rv{Edge detection and image saliency} are integrated to identify unimportant regions for text label placement.
Jia~et~al.~\cite{imagebased} investigated a similar strategy, \rv{with incorporating human placement preferences as a set of constraints} to improve the work by Grasset~et~al.~\cite{grassetimage}.
\rv{Two prototypes are implemented for desktop computers due to the poor \rv{temporal performance} on mobile devices}.
Tatzgern~et~al.~\cite{hedgehog} developed a pioneering approach that considers labels as 3D objects in the scene to avoid unstable labels due to view changes.
The approach estimates the center position of an object and moves labels along a 3D pole, which attaches to the object. Another proposed scenario constrains label movement to a predefined 2D view plane. 
\rv{This technique is limited to} annotating \rv{objects} in front of the camera.

\rv{Existing work tends to directly solve label occlusions in 2D or \rv{to} project labels from 3D to 2D and apply 2D solutions.
\rv{These techniques cannot avoid label inconsistencies}~\cite{Cmolik:TVCG:2020}.
In contrast to existing approaches, we handle labels as objects in the 3D scene. This allows us to compensate for incoherent label movement caused by viewing angle changes of the device.} 
\rv{We integrate} the labeling technique \rv{into} 3D to retain stability and introduce additional visual variables, including text, images, icons, and colors, to enrich the corresponding visual representation.
Our label encoding also varies in order to balance information provided by POIs.
More design choices will be explained in Section~\ref{sec:overview}.

%% file: sections/overview.tex
\section{Context-Responsive Framework}
\label{sec:overview}

\rv{Based on the taxonomy by Wiener et al.~\cite{Wiener:2009:SCC}, our approach supports aided and unaided wayfinding tasks. We can directly highlight the destination label and assist users to combine decision-making processes, memory processes, learning processes, and planning processes for finding the overall best destinations.} The effort to identify objects in AR is low~\cite{guarese} because real-world objects can be directly annotated~\cite{firstnavigation, grassetimage} and AR navigation \rv{is less user-focus demanding} compared to \rv{other map techniques~\cite{McMahon:2015:JSET}.} 
The responsive framework is inspired by Hoffswell et al.~\cite{Hoffswell:2020:CHI}, who proposed a taxonomy for responsive visualization design, which is essential to present information based on the device context.
In principle, our design has three major components, including (1) occlusion management, (2) level-of-detail management, and (3) coherence management, each of which aims to solve the problems \textbf{(P1-P3)}, respectively. 
We first introduce the encoding of labels \rv{beyond plain text,} followed by an overview of the \rv{presented} approach.
% \begin{itemize}
%     \item The effort to identify an object is lower~\cite{firstnavigation, imagebased, guarese}.
%     \item Real-world objects can be directly annotated~\cite{firstnavigation, grassetimage}.
%     \item AR navigation requires less user focus demand~\cite{guarese}.
%     % \item There is no need to translate navigation from a 2D view, which makes navigation clearer and more intuitive \cite{guarese}. 
% \end{itemize}
%

%Due to the intuitiveness of using AR in navigation and its challenging when annotating POIs, we aim for a better solution in this project.

% \subsection{Overview and Design Principles of Our Responsive Framework}
% \label{ssec:design}

\subsection{Label Encoding}
\label{ssec:encoding}

The label encoding \rv{reduces the limitations in existing work and solves \textbf{(P2)}. We introduce additional types of labels than merely text labels as concluded by Langlotz~et~al.~\cite{nextgen}}. 
% The labels depicted directly in the AR domain solve problem. 
We use color to encode \rv{scalar variables of each POI~\cite{suitablecomp,mazza}.}
In general, the users can choose a color scheme and a scale according to their preferences.
A label consists of several of the following components: 
\begin{itemize}
    \item a text tag containing the name of the POI, 
    \item an iconic image (photo) of the POI, 
    \item an icon encoding the type of the POI, and
    \item a color-coded rectangle representing a scalar value of the POI.
\end{itemize}

\begin{figure}[tbh!]
\centering
\setlength{\tabcolsep}{1pt}
\begin{tabular}{cc}
\includegraphics[width=0.48\linewidth]{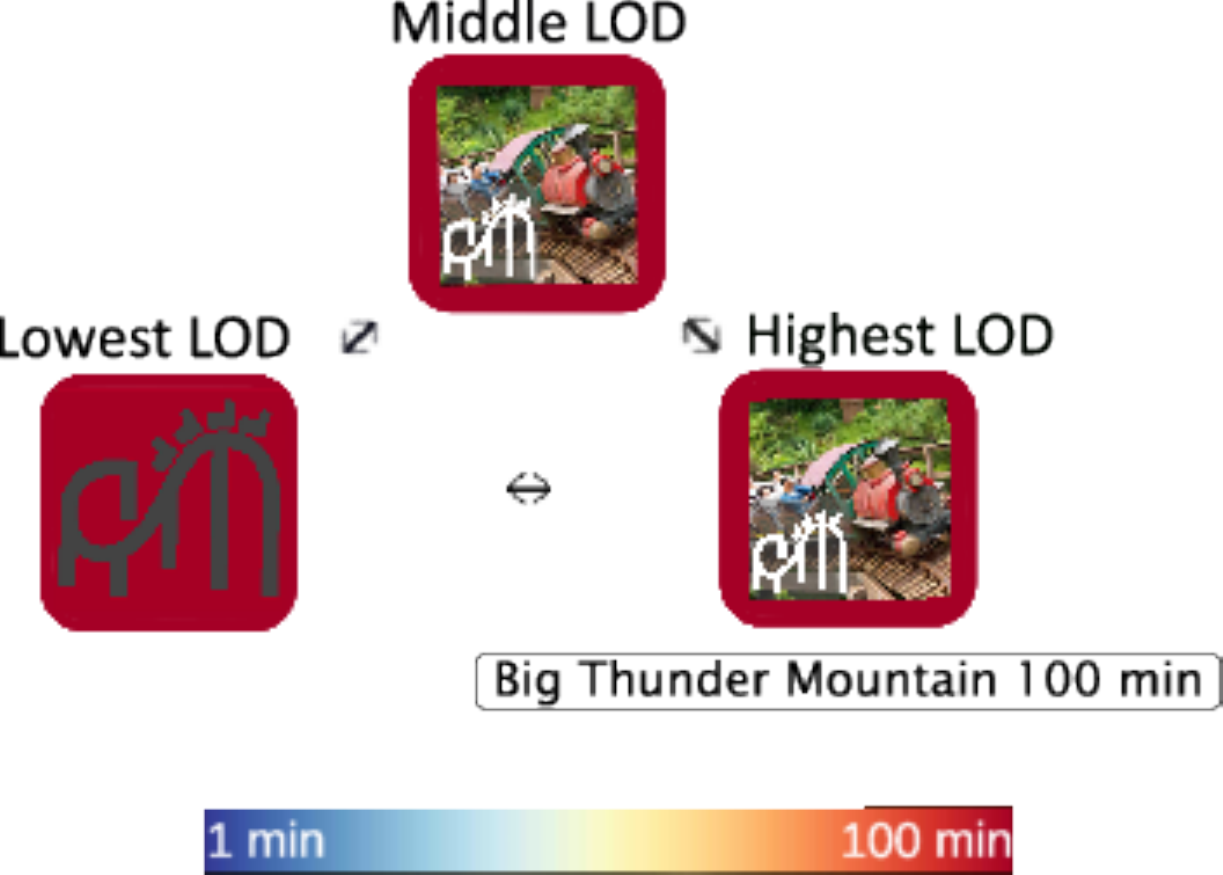} &
\includegraphics[width=0.48\linewidth]{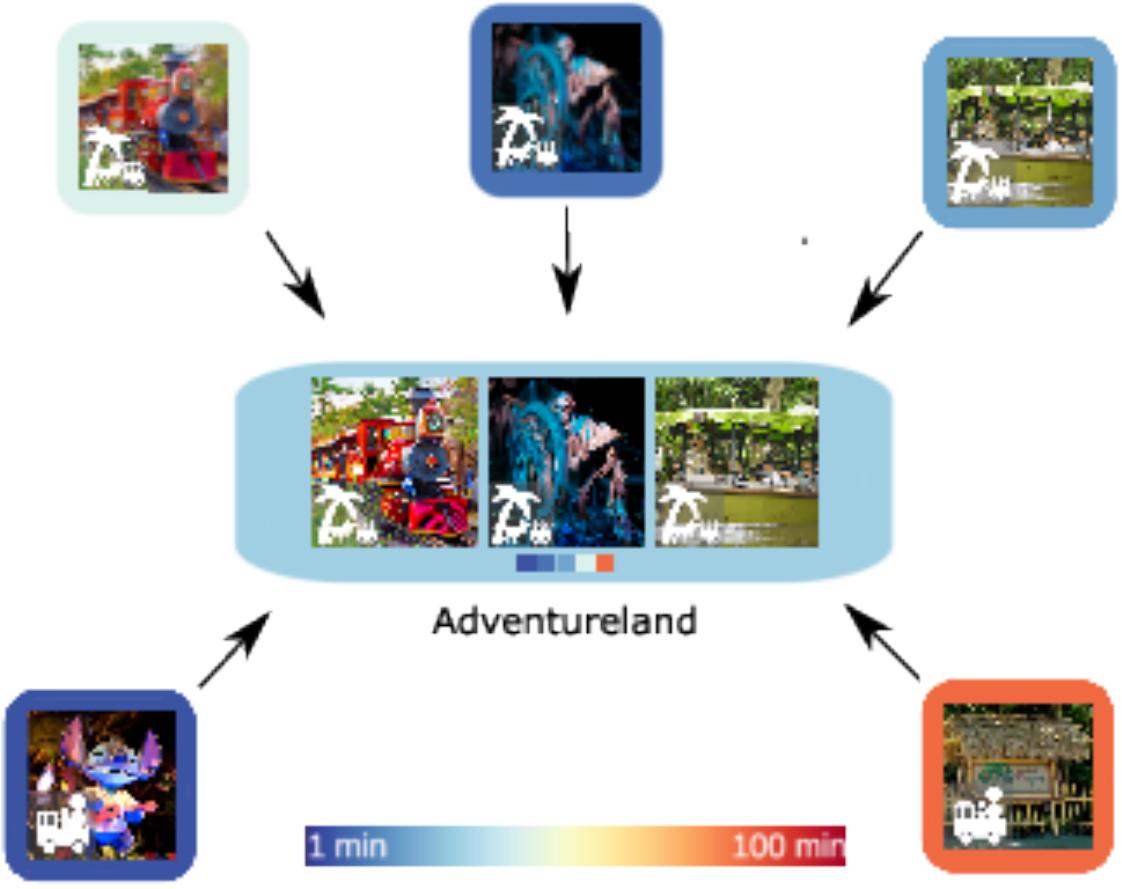} \\
\rv{(a) Label encoding, three LODs} & \rv{(b) Super label} \\
\end{tabular}
\caption{An example label encoding \rv{(\emph{Tokyo Disneyland Dataset})}.}
\label{fig:lodsfirst}
\end{figure}

\begin{figure*}[htb!]
\centering{
 \setlength{\tabcolsep}{1pt}
 %\begin{tabular}{ccccc}
 \begin{tabular}{ccccc}
   %%%%%%%%%%
    \includegraphics[width=0.15\linewidth]{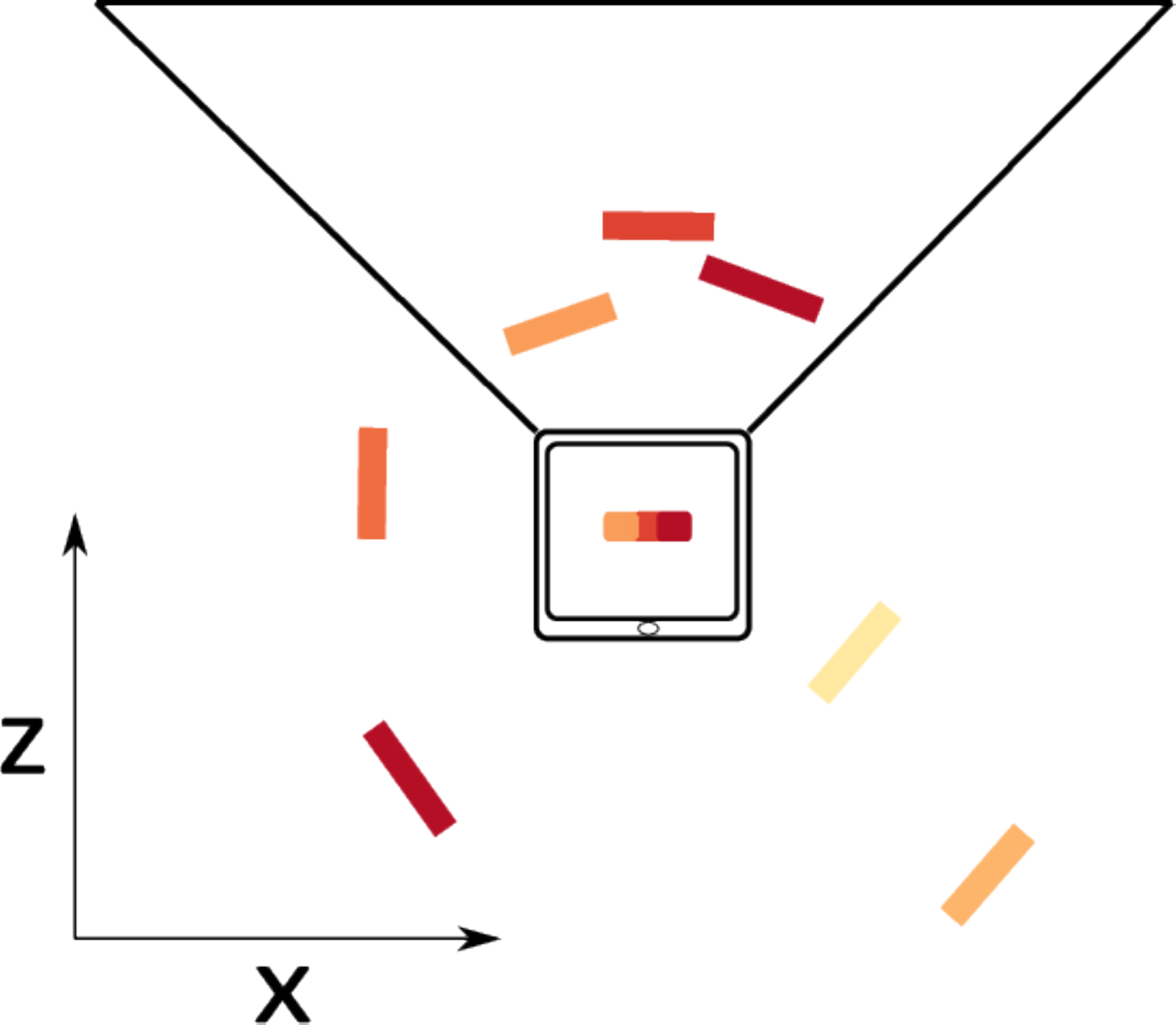} &
    \includegraphics[width=0.15\linewidth]{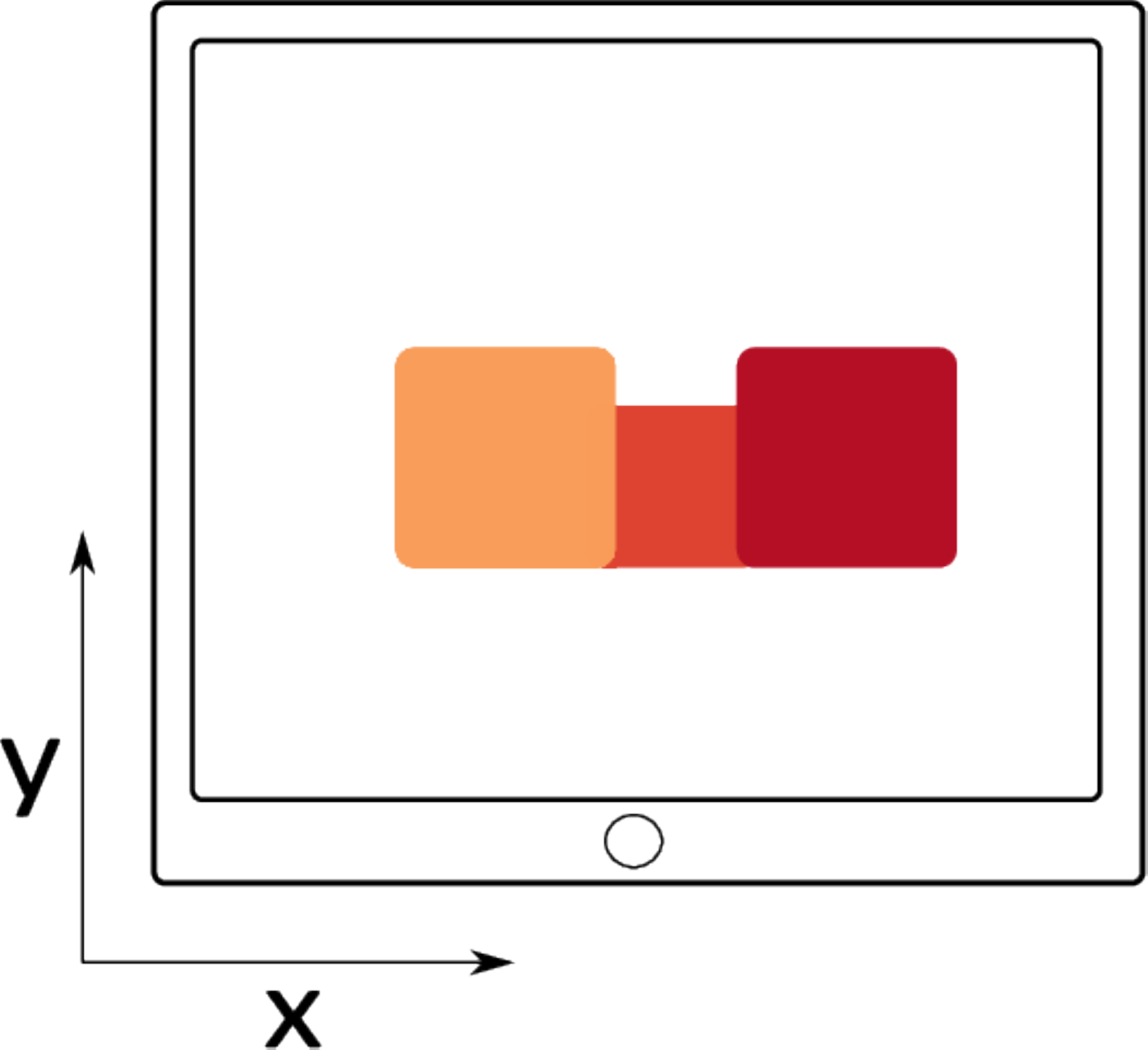} &
    \includegraphics[width=0.15\linewidth]{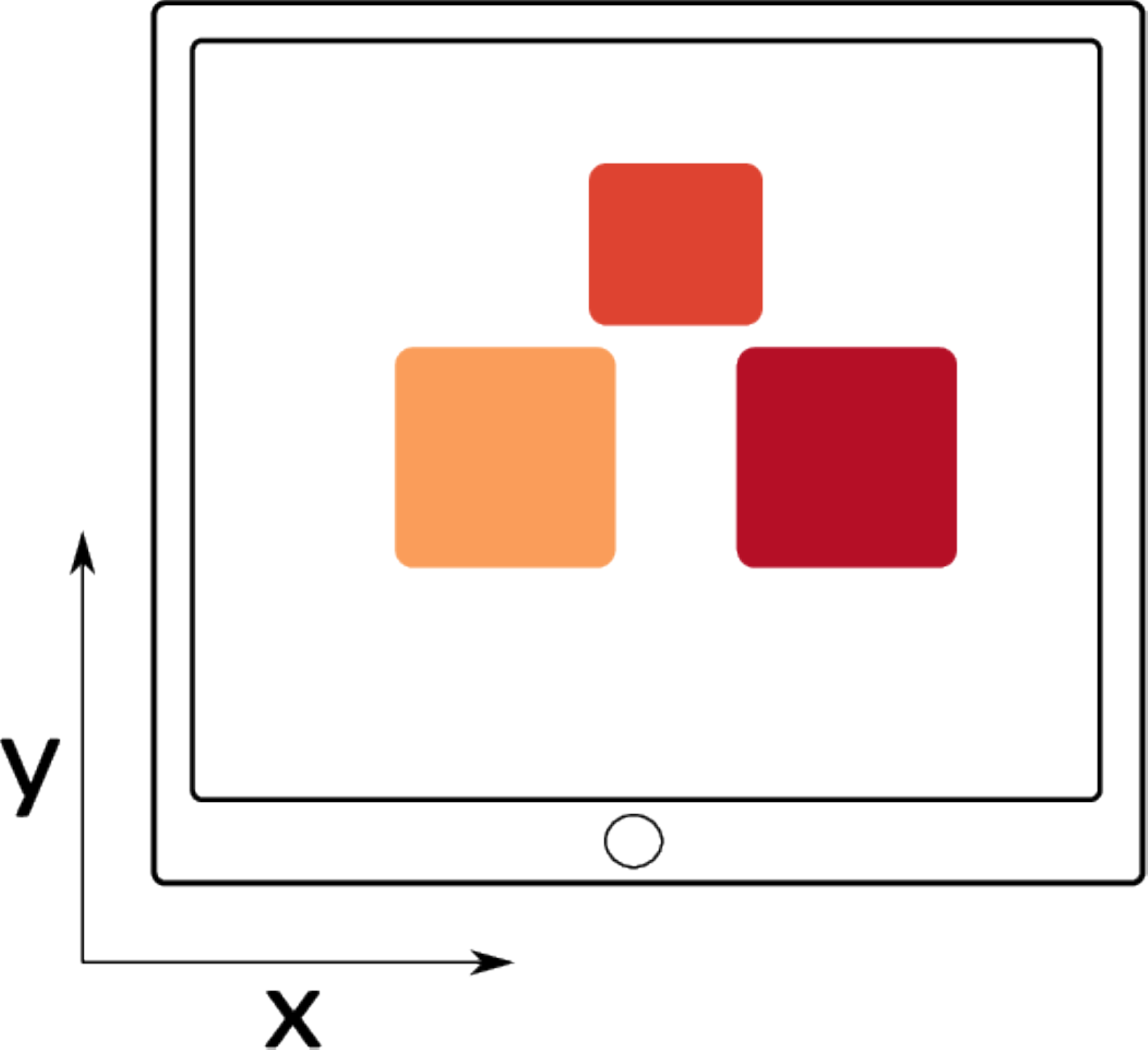} &
    \includegraphics[width=0.15\linewidth]{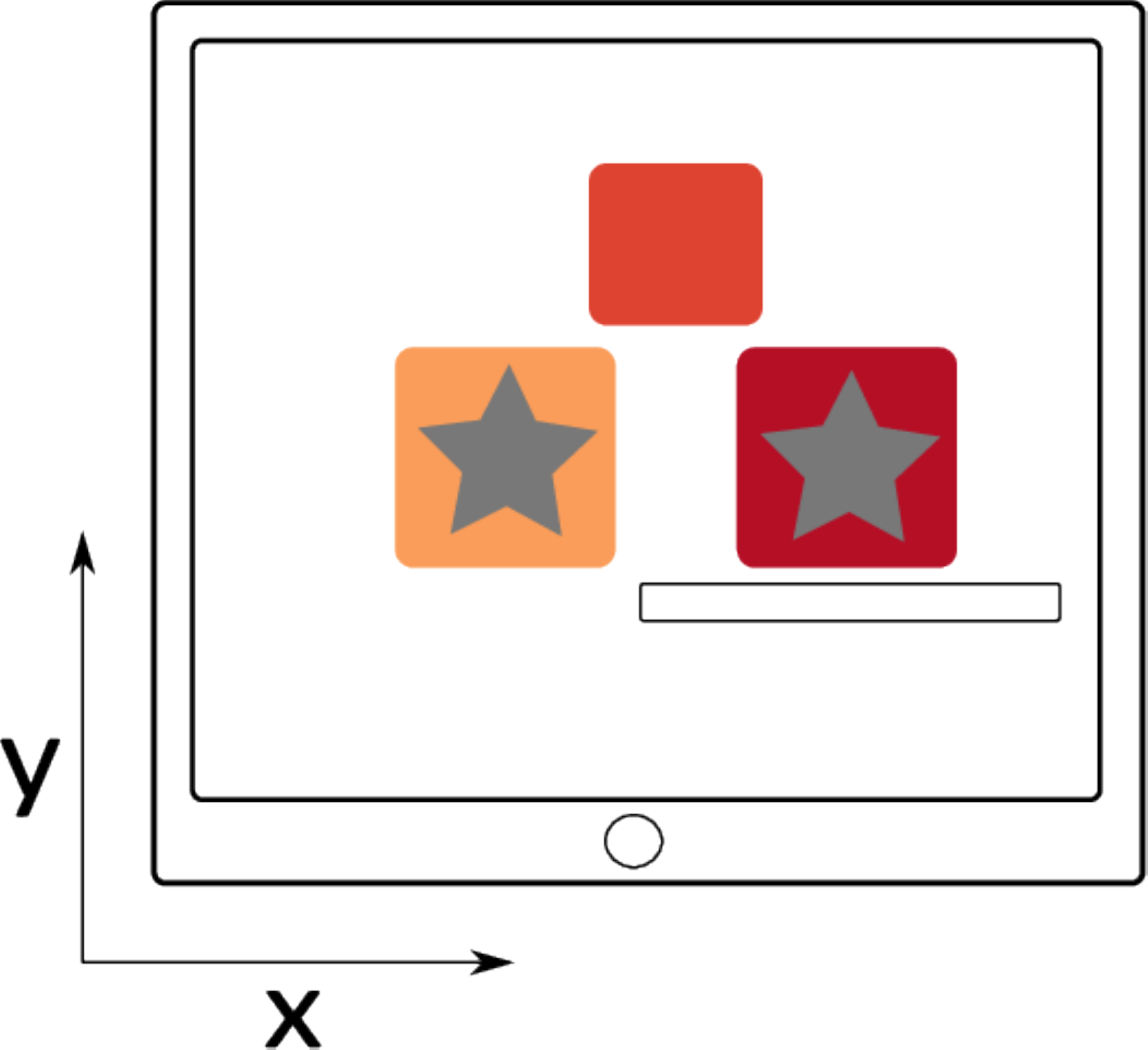} &
    \includegraphics[width=0.15\linewidth]{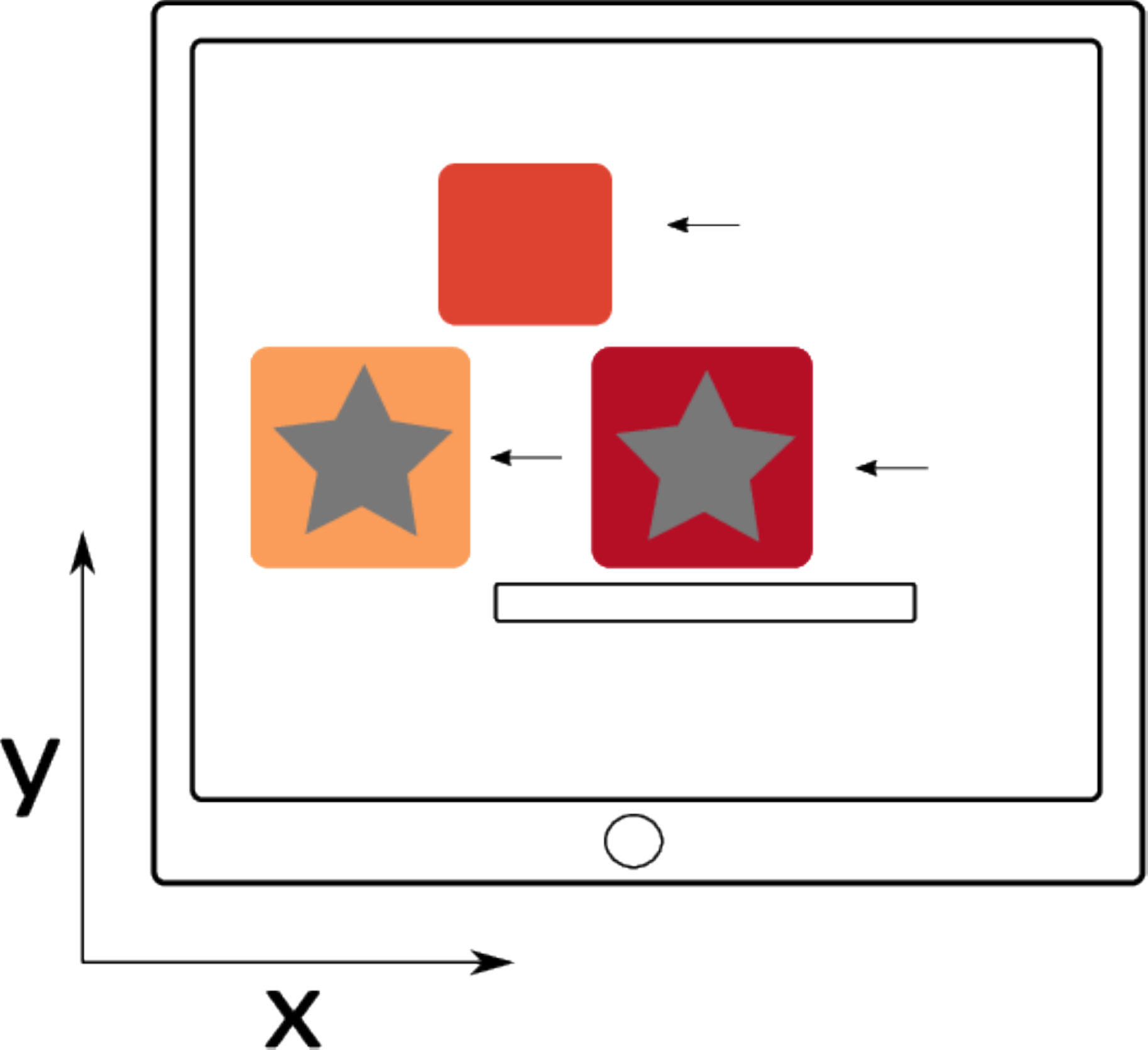} \\
   %%%%%%%%%%
   (a) Input & 
   (b) Positioning labels in AR & 
   (c) Occlusion management& 
   (d) Level-of-detail management& 
   (e) Coherence management \\ 
 \end{tabular}
}
\caption{The input scenario (a), positioning of labels in AR (b), and the three management strategies of our approach (c)-(e).}
 \label{fig:overview}
\end{figure*}

\rv{In Figure~\ref{fig:lodsfirst}, labels concerning the \emph{Tokyo Disneyland Dataset} are shown.} 
POIs are attractions in this case. 
\rv{Attractions can be categorized into three types, i.e., \emph{thrilling}, \emph{adventure}, and \emph{children}, each of which is depicted through a type icon.} 
Figure~\ref{fig:lodsfirst}(a) provides an explanatory label annotating an attraction of the dataset. 
The text tag depicts the name of the attraction and the waiting time (e.g., \emph{Big Thunder Mountain $100$ min}). The iconic image shows a photo of the train of the attraction and the type icon indicates that it is a thrilling attraction. The colored \rv{(rectangular) backgrounds of the labels encode} the corresponding waiting times. 

\subsection{\rv{Pipeline} of the Approach}

Figure~\ref{fig:overview} gives an overview of our approach. 
We first position labels of POIs in AR (Figure~\ref{fig:overview}(a) as a top view and (b) as a front view) and perform the proposed three management strategies.
% After initially positioning the labels in AR (Figure~\ref{fig:overview}(a) as a top view and (b) as a front view), we can step by step follow the main components in our responsive framework.
% consists of three management strategies, including (1) \emph{occlusion management}, (2) \emph{level-of-detail management}, and (3) \emph{coherence management}. 
 %The labels are initially positioned according to their real-world location. Our 3D coordinate system encodes the latitude and longitude values of each POI along the $x$ and $z$ axes. This location based positioning is important for the navigation process \cite{firstnavigation, guarese}.
We \rv{process the objects} in the 3D scene using a Cartesian world coordinate system, where the $xz$-plane is parallel to the ground plane.
Figure~\ref{fig:overview}(a) depicts a top view of our coordinate system, the $x$-axis and $z$-axis define the ground plane and the $y$-axis is vertically upwards from the ground plane. The input to our system is a set of POIs $P = \{p_1,p_2,...,p_n\}$ and a set of labels $L = \{l_1,l_2,...,l_n\}$, for example, manually selected by the users or downloaded from an online database. In the \emph{positioning labels in AR} preprocessing (Section~\ref{ssec::arpos}), for each POI $p_i$, the corresponding label $l_i$ is initially \rv{placed} perpendicularly to the ground plane in the world coordinate system (Figure~\ref{fig:overview}(b)). Currently, each POI $p_i$ has one associated label $l_i$ describing the attributes of the POI. We also assume that the $(x, z)$-coordinates of each annotated POI are more important than the $y$-coordinate, since the $(x, z)$-coordinates are essential to indicate the relative positions of the POIs \rv{as suggested by prior work~\cite{firstnavigation, guarese}.}

The \emph{occlusion management} strategy (Section~\ref{ssec:occclusion}) \rv{addresses} \textbf{(P1)} and resolves occlusions of labels considering the current configuration of the device. 
The labels are first sorted by distance from the device \rv{into} a list $S$, from the nearest to the farthest positions. With this information, we resolve occlusions starting with the closest label and using a greedy approach (see Figure~\ref{fig:overview}(c)). The greedy approach arranges the lowest $y$-positions of the labels to be visible iteratively. This allows effective execution of the occlusion-handling on mobile devices, where the computation powers are limited compared to desktop computers. The occlusion strategy provides a solution to \rv{otherwise} inconsistently moving labels when the viewing angle of the AR device changes~\cite{hedgehog}. 
% Our \emph{occlusion management} (Section~\ref{ssec:occclusion}) solves occlusions considering the users' position for each label $l_i$ in the world space. 

%The closest label stays at its initial position. 
%We then iteratively check, if occlusions happen in 3D in the current scene and eliminate them accordingly (Section~\ref{ssec:occclusion}). Due to the importance of annotating POIs at the respective $(x, z)$-coordinates, we shift labels upwards along the $y$-axis to resolve occlusions. We do not change the $(x, z)$-coordinates of labels to keep necessary computation times on mobile devices low, which is preferred by the users.
%We apply a greedy algorithm as an optimal placement is NP-hard \cite{nphard} in 2D. Our occlusion strategy provides a solution to inconsistently moving labels when the viewing angle of the AR device changes as stated by Tatzgern~et~al.~\cite{hedgehog}. Our \emph{occlusion management} (Section~\ref{ssec:occclusion}) solves occlusions considering the users' position for each label $l_i$ in the world space. 

In the \emph{level-of-detail management} (Section~\ref{ssec:module2}), we introduce four distinct types of label encodings for \textbf{(P2)} to represent three levels-of-detail (LODs, Figure~\ref{fig:lodsfirst}(a)) of an individual label and one \emph{super label} to indicate an aggregated group of labels for visual clutter reduction (Figure~\ref{fig:lodsfirst}(b)).
\rv{The \emph{level-of-detail management} depicts} a different amount of information for each label (see Figure~\ref{fig:overview}(d)). The LOD of a label $l_i$ is selected according to the distance of the annotated POI to the device and the label density in the view volume. For convenience, we assume that \rv{close labels get at least as much screen space as distant labels}, since it is natural to show objects larger when they are close by. \rv{However, different configurations can be also incorporated by adding rays in the occlusion detection.} %For this reason, close labels in our system provide more detailed information compared to labels that are located in the distance. 
%This decision is made based on the assumption that closer labels are \rv{either larger than or as large as distant labels, since it is natural that close objects are larger}.
%
Super labels (Figure~\ref{fig:lodsfirst}(b)) are representative labels that \rv{depict a set of aggregated labels in order to reduce visual clutter. Figure~\ref{fig:lodsfirst}(b) gives} an example of a super label for the \emph{Tokyo Disneyland Dataset}. 
The themed area \emph{Adventureland} is aggregated and \rv{the blue background color} of the super label encodes the average waiting time. 
\rv{A color legend at the bottom of the label presents the individual waiting times of the aggregated attractions in this themed area.} 

\emph{Positioning labels in AR}, \emph{occlusion management}, and \emph{level-of-detail management} are smoothly updated in the \emph{coherence management} module (Figure~\ref{fig:overview}(e)). To avoid flickering that inevitably reduces coherency~\cite{imagebased}, the labels are not moved or changed immediately, but follow a common animation policy, by strategically updating changes over time (Section~\ref{ssec:coherencemanagement}) to solve problem \textbf{(P3)}.

%% file: sections/arsetting.tex
\section{Context-Responsive Labeling Management}
\label{sec:method}

Our approach positions labels in AR space, followed by a context-responsive computation. Here we introduce occlusion removal, perform level-of-detail strategies, and enforce coherent label placement. In this section, we will detail the proposed technique.

\subsection{Positioning Labels in AR}
\label{ssec::arpos}

In \rv{a} preprocessing step, \rv{we map} the geographical locations from the real world to our Cartesian AR world space. 
% Real-world POIs can be mapped to the AR world space 
\rv{This considers} the GPS position of the user's device, the GPS location of the POIs, and the compass orientation of the device~\cite{gpsPositioning}.%\cite{Maass:2006:WSCG}. 

The labels are oriented \rv{towards} the user's position by aligning the normal vectors of the labels with the AR device in the AR world space. 
Once this initial label positioning is done, a perspective projection from the AR world space into the screen space of the device is performed. \rv{In doing so, we can position the labels in AR spatially relative to the position of the user to support exploration and navigation as shown by Guarese~et~al.~\cite{guarese}.}
In principle, existing frameworks, like the \emph{AR + GPS Location SDK} package~\cite{gpslocation} or the \emph{Wikitude AR SDK} package~\cite{wikitude} can be used to map real-world objects to the AR world space. 
Unfortunately in our experiment, the techniques are not stable \rv{due to} the inaccurate GPS sensor~\cite{lowCost} or compass~\cite{kuhlmann} \rv{data} of mobile devices.
% Moreover, the documentations also do not give an insight into how exactly this mapping is calculated.
To test and assess the quality of the coherence strategies for the \emph{occlusion management} and \emph{level-of-detail management}, \rv{we predefine the positions of labels} at the $(x, z)$-coordinates in the AR world space.
The existing libraries do not provide stable label positions, which would lead to a less coherent behavior that is not relying on the proposed \emph{coherence management}. Once the labels are placed, 
we order the labels based on their distance to the user for future computations.

%% file: sections/occlusion.tex
\begin{figure*}[tbh!]
\centering{
 \setlength{\tabcolsep}{1pt}
 \begin{tabular}{cccc}
   %%%%%%%%%%
    \includegraphics[width=0.17\linewidth]{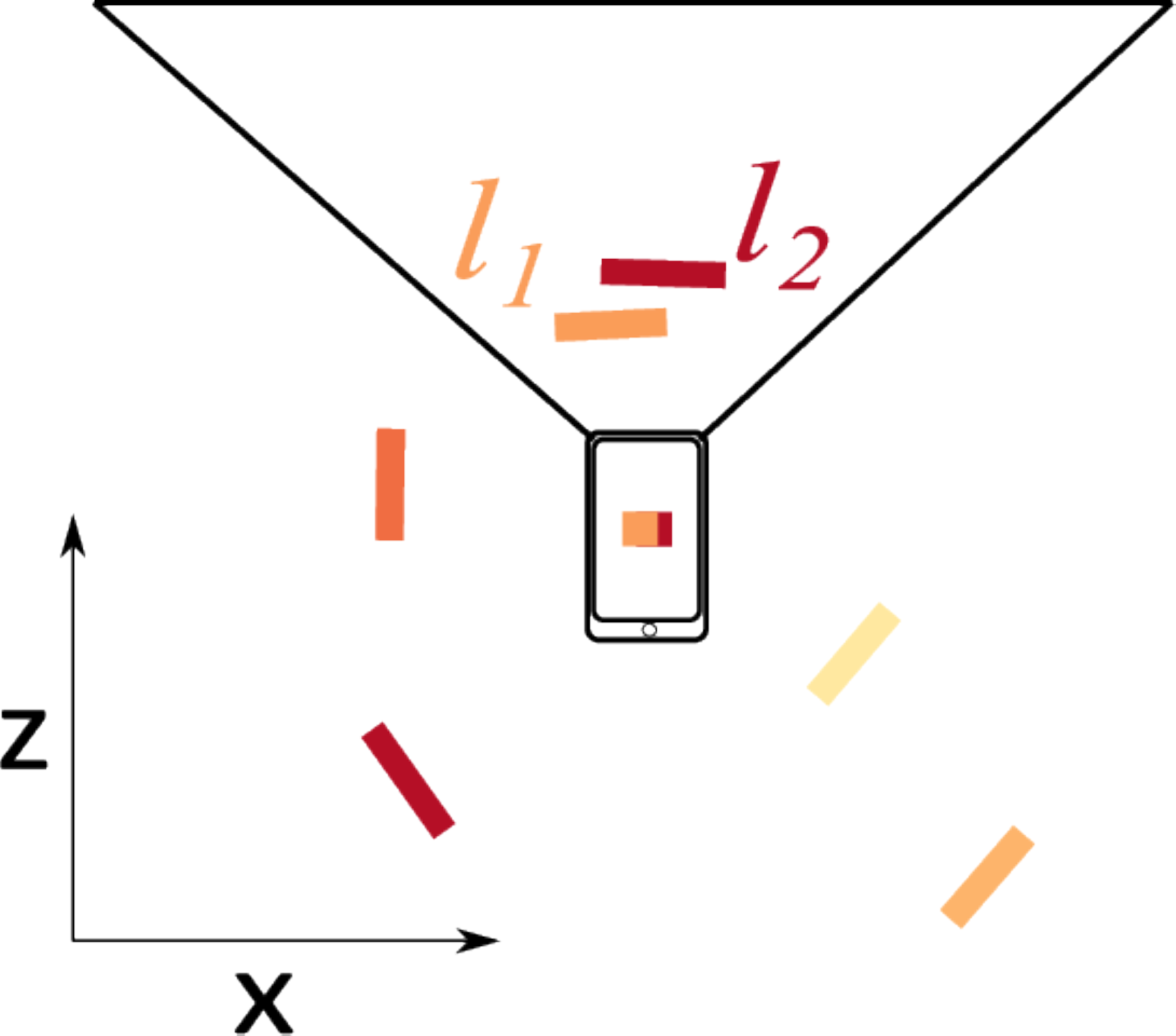} & 
   % \includegraphics[width=0.19\linewidth]{images/method/FourPointsRayCasting2.eps} & 
    % \label{fig:fourpoint} 
    \includegraphics[width=0.17\linewidth]{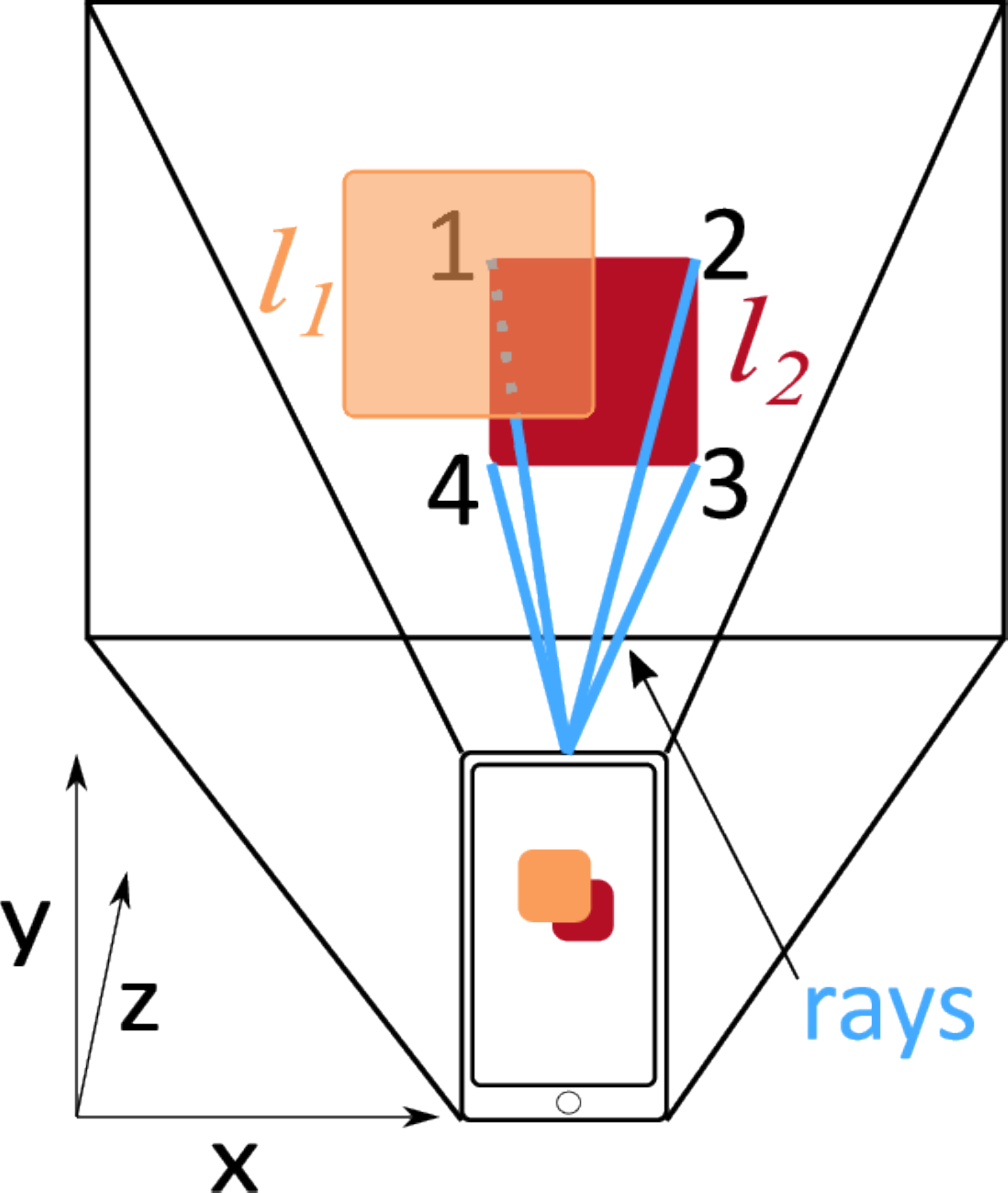} &
    \includegraphics[width=0.27\linewidth]{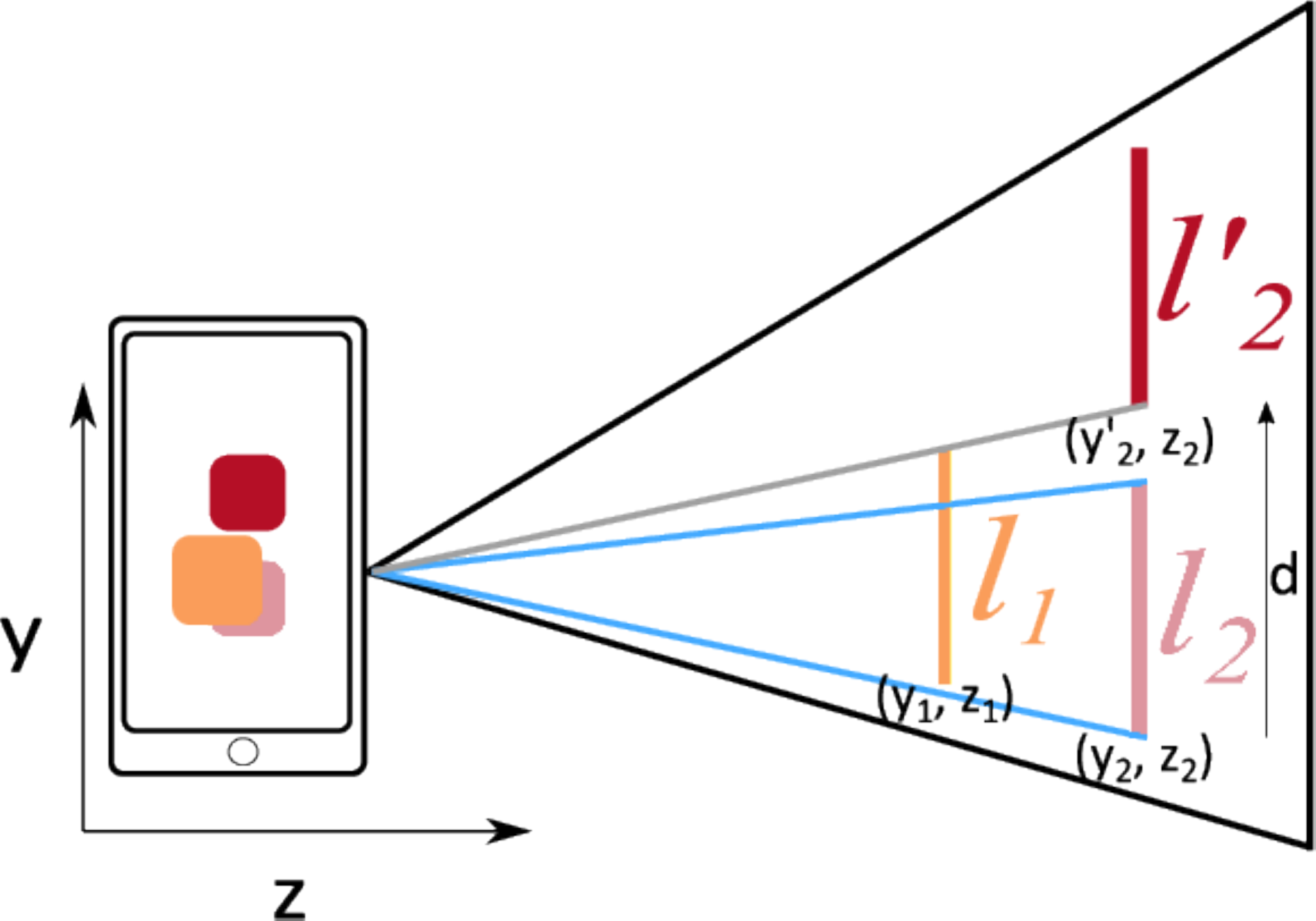} &
    \includegraphics[width=0.27\linewidth]{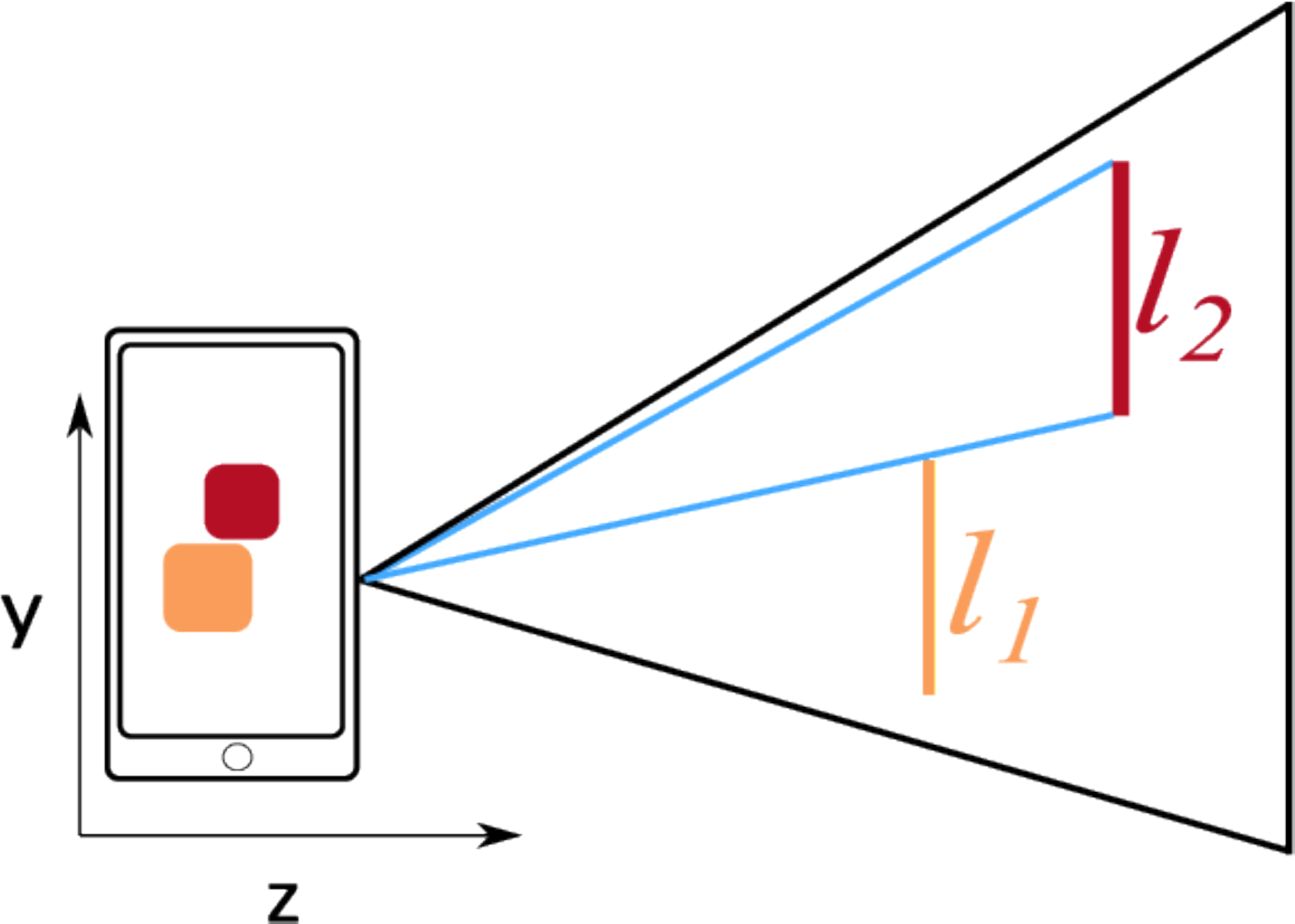}  \\
   %%%%%%%%%%
   (a) World coordinates (top view)&
   %(b) four point \label{fig:fourpoint} &
   (b) Occluded label $l_2$ \label{fig:detect} & 
   (c) Shift of $l_2$\label{fig:shiftComp} & 
   (d) Occlusion-free result \label{fig:finalocclusion} \\ 
 \end{tabular}
}
\caption{
\rv{Illustration} of the \protect\emph{occlusion management}.
(a) The labeled 3D scenario in top view. Label $l_1$ and label $l_2$ are in the current view volume.
(b) Occlusion by $l_1$ (transparent and orange), which is in front of $l_2$ (red). The ray at corner $1$ of $l_2$ intersects the occluding, label $l_1$. 
(c) Label $l_2$ is shifted above the gray ray by the distance $d$ to resolve the occlusion.
(d) The blue corner rays do not collide with a label in front anymore. 
}
 \label{fig:manage}
\end{figure*}

\subsection{Occlusion Management}
\label{ssec:occclusion}

Showing many labels simultaneously on a mobile device will, unfortunately, lead to occlusions of labels, especially if the annotated POIs are close to each other or even hidden by other \rv{labels (Figure~\ref{fig:manage}(a)).} 
Point-feature labeling has been extensively investigated due to its NP-hardness, even when looking for an optimal solution just in 2D~\cite{nphard}. In our setting, occlusions change over time, since the users move. Fast responsive management strategies are required to update the scene regularly. Viewing angle and position changes of the user need to be accounted for to guarantee smooth state transitions and to eliminate unwanted flickering. We \rv{perform} the entire occlusion handling in the 3D scene, overcoming the label positioning inconsistencies caused by viewing angle changes. The occlusion handling consists of two steps, occlusion detection and shift computation.

\subsubsection{Occlusion Detection}

%\cite{grassetimage, imagebased, hedgehog, labelsurvey}
\rv{We employ ray tracing to detect occlusions, which is different from existing approaches~\cite{labelsurvey}}. As the labels have been sorted by the distance to the user, the occlusions are detected and solved iteratively from label $l_1$ to label $l_n$ of the sorted list $S$. For each label $l_i$, the origins of four rays are set to the location of the user's device in AR. \rv{The rays run through the corner points} of \rv{label $l_i$ as shown in Figure~\ref{fig:manage}(b).} 
If another label is hit during the ray traversals, \rv{an occlusion occurs.} To ensure that all possible occlusions will be detected, we assume that \rv{labels} closer to the viewer are either larger or as large as labels \rv{farther away}. This allows us to \rv{use just} four rays to detect 3D occlusions effectively. 
The approach works for rectangular shapes or rectangular bounding boxes of polygonal shapes and could be extended to polygons \rv{or 3D objects (e.g., buildings in MR)}.
\rv{Other configurations can be accommodated by increasing the number of rays.}
Figure~\ref{fig:manage}(b) gives an example,
where label $l_1$ (orange) is in front of label $l_2$ (red). In this case, the corner ray $1$ of label $l_2$ collides with label $l_1$, indicating that label $l_1$ occludes label $l_2$.  Since we assume that closer labels are always larger or as large as \rv{farther away labels}, no \rv{occluding} labels will be missed during the occlusion detection.

\subsubsection{Shift Computation}

\rv{Once the occlusions are detected, we can iteratively shift the labels greedily in the order of increasing distance.}
%
% \begin{algorithm}[h]
%     \SetAlgoLined\KwData{Labels}
%     \KwResult{Occlusion-free labels}
%     $S$ = list of sorted labels\;
%     \For{$l_i \in S$}{
%         create four rays to corners of $l_i$\;
%         \While{ray collides with label}
%         {
%             get occluding label\;
%             shoot ray from user to top of occluding label\;
%             traverse ray to $(x, z)$ position of $l_i$\;
%             shift $l_i$ above this point\;
%         }
%         {//$l_i$ is completely visible\;}}
%     \caption{Simplified occlusion handling through shifting.}
%     \label{algo:basic}
% \end{algorithm}
%
Since the labels are shifted from the closest to the farthest one, the label $l_i$ will be located either at its initial $(x,z)$-coordinates or above the previous label $l_{i -1}$ along the y-axis. 
Figure~\ref{fig:manage}(c) illustrates the basic shift of label $l_2$. The blue lines represent the \rv{corner rays for occlusion detection} and the gray line shows the traversed ray for calculating the occlusion free position of label $l_2$. Figure~\ref{fig:manage}(d) depicts an occlusion-free result after \rv{shifting} label $l_2$, where the shift distance $d$ is $|y_2'-y_2|$.

\rv{Szirmay-Kalos et al.~\cite{worstcase} proved that the ray-tracing approach at least requires a logarithmic computation time in the worst case based on the number of scene objects. On the other hand, modern platforms already provide real-time ray-tracing~\cite{unity}.}
In our approach, the \emph{occlusion management} takes $O(n^2)$ if labels are aligned in a sequence \rv{along} the current viewing direction.
The current label $l_i$ \rv{possibly} needs to be shifted above each label in front of it. We show a comparison with different label alignments in Section~\ref{sec:result}. \rv{The greedy label placement terminates} as soon as no other label in front occludes label $l_i$.

%% file: sections/lod.tex
\subsection{Level-Of-Detail Management}
\label{ssec:module2}

\rv{Labels occupy space that is a scarce resource on a mobile device}, especially if many labels should be shown simultaneously. 
To reduce \rv{unwanted visual clutter, we introduce an} LOD concept for labels~\cite{matkovic} and incorporate a \emph{level-of-detail management} in the pipeline (Figure~\ref{fig:overview}(d)). The LOD is also computed based on the \rv{sorted distances of labels} and the label density.% in the view volume aligned along the $(x, z)$ horizontal ground plane. 
The LOD selection consists of two steps: LOD calculation and super label aggregation.

%The system includes three LODs for each individual label and super labels that aggregate \rv{a group of} labels as presented in Figure~\ref{fig:lodsfirst}. 
In our implementation, the \rv{lowest} LOD occupies the least space and includes a colored rectangle and an icon (Figure~\ref{fig:lodsfirst}(a)). The middle LOD presents a colored rectangle, the icon, and an iconic image (photo) of the POI (Figure~\ref{fig:lodsfirst}(a)). The \rv{highest LOD contains} a text tag and occupies the most space (Figure~\ref{fig:lodsfirst}(a)). The level-of-detail for each label changes when the user navigates through the scene.

\begin{comment}
\begin{figure}[h]
\centering
  \includegraphics[width=0.8\linewidth]{images/method/LODs.eps}
  \caption{Our three LODs for each label. The arrows between them indicate the possible state changes over time.}
  \label{fig:lodsfirst}
\end{figure}
\end{comment}

\subsubsection{LOD Calculation}

The LOD for each label depends on the distance to the user and the label density. For each label, a virtual view volume aligned to the $(x, z)$ ground plane is constructed to mimic that the user would look into the direction of each label. 
The horizontal distance along the $(x, z)$ ground plane and the vector from the user to the position of each label are used. If the angle between these two vectors is above a threshold $t$ ($45^{\degree}$ by default in our system), the label is located outside the aligned view volume. 
\rv{We} split the view volume and each label below the threshold $m_1$ ($20^{\degree}$ by default) receives the \rv{highest} LOD until one label \rv{exceeds} the angle $m_1$. 
\rv{The remaining labels are displayed in the middle LOD until reaching the threshold} $m_2$ ($30^{\degree}$ by default). 
\rv{If a label exceeds $m_2$,} it will be displayed in the \rv{lowest} LOD. \rv{The threshold angles can} be changed according to user preferences. 
The LODs of all labels are consistent when the viewing angle of the device changes for the current user position. 
The \emph{level-of-detail management} provides coherent movement when rotating the AR device. The LODs for the labels are updated if the user \rv{moves.} 

\subsubsection{Super Label Aggregation}

To further reduce visual clutter, we introduce super labels \rv{that group individual labels} \rv{(see Section~\ref{ssec:encoding})}. The position of a super label is calculated as the average $(x,z)$-positions of the individual labels that are part of the aggregation in the 3D scene. A predefined grouping \rv{(i.e., themed areas of amusement parks)} of labels is necessary to compute the super labels, \rv{while unsupervised clustering algorithms can also be directly applied.} We do not aggregate labels of the closest predefined group considering the position of the user. Individual labels in the close surroundings of the user are always displayed and not aggregated supporting the exploration process. \rv{We} only aggregate individual labels to super labels if the user is located outside of the \rv{respective label group.}

%% file: sections/transition.tex
\subsection{Coherence Management}
\label{ssec:coherencemanagement}

To avoid unwanted flickering, we incorporate smooth transitions for each movement and change. \rv{Smooth transitions are implemented} if positions of labels change to be occlusion-free during the interaction with the system, if LODs of labels change, or if labels are aggregated to super labels. \rv{We investigated ten different easing functions, including linear, and various quadratic and cubic equations, for the transitions to further increase the coherency. 
\rv{For comparison, we refer readers to the supplementary videos.}
We believe that the ease-in ease-out sine function (Eq.~\ref{eq:completion}) represents the best easing function as it provides harmonic transitions. The easing function can be changed based on user preferences. Let $t_{transition}$ be the duration for \rv{a} transition to be completed. The variables $t_{start}$ and $t_{current}$ indicate the start time and the current time \rv{during} the transition. The function $e(t_{current})$ represents \rv{the} easing function for\rv{ a smooth transition}:}
\rv{
\begin{eqnarray}
e(t_{current}) = -0.5 * (\cos{(\pi * \frac{t_{current} - t_{start}}{t_{transition}}}) - 1).
\label{eq:completion}
\end{eqnarray}
}

\subsubsection{Smooth Occlusion Transitions}

Due to the interaction of the user, occlusion-free label positions may vary from one frame to the next. If the labels would simply be displayed at the newly calculated positions, the labels \rv{might} abruptly change their positions, which destroys the users' experience since the labels do not move in a coherent way. \rv{To allow the user to better keep track of the labels, we implemented smooth transitions from the previous locations of the labels to the newly calculated ones. \rv{We interpolate }original positions and the newly calculated positions of the labels. 
The position for label ${l_i}$ is updated every frame until it reaches its destination. Let $p_{goal}({l_i})$ be the new occlusion-free label position and $p_{start}({l_i})$ the label position at the start of the transition. We calculate the current label position for label ${l_i}$:}
% as shown in Equation~\ref{eq:occlTrans}.}
%
%
\begin{eqnarray}
\rv{\vec{p}(l_{i}) = \vec{p}_{start}({l_i}) + (\vec{p}_{goal}({l_i}) - \vec{p}_{start}({l_i})) * e(t_{current})}.
\label{eq:occlTrans}
\end{eqnarray}

\subsubsection{Smooth LOD Transitions}

If the LOD for a label changes, the transition needs to be smoothed to avoid flickering and allow a coherent user experience. The LODs of labels change over time, and we adapt the alpha channel to achieve a smooth transition. In this way, the iconic images, the icons, and the text tags fade in or out using
\rv{
\begin{eqnarray}
    \alpha(l_{i})=\begin{cases}
                          e(t_{current}), & b = 1 \\
                          1 - e(t_{current}), & b = 0,
                       \end{cases}
                       \label{eq:lodtrans}
\end{eqnarray}
where} $\alpha(l_{i})$ is the alpha value of the iconic image, the icon, or the text tag of label $l_{i}$. \rv{Since our easing function $e$ (in Eq.(\ref{eq:completion})) returns a value between $0$ and $1$, the result can be used to set the alpha channel in Eq.(\ref{eq:lodtrans})}. The variable $b$ indicates, if the object should become invisible ($b=0$) or if the object should become visible ($b=1$). 

\subsubsection{Smooth Aggregation Transitions}

\rv{If} labels are aggregated to super labels, individual labels will be moved to the respective super label positions in the scene. 
Simultaneously, we fade in the super labels and fade out the labels by interpolating the alpha channels. \rv{If} individual labels are aggregated, the labels move towards their super label and disappear. If an aggregation is split up again, coherency is achieved analogously. 
If the alpha channel of a super label is decreased, the individual labels reappear over time and \rv{move} back to their respective positions (Eqs.(\ref{eq:superal}), (\ref{eq:superas}), and (\ref{eq:superpos})). Let $l_i$ be a label that will be aggregated into \rv{a} super label $l_{s}$.
\rv{The alpha values of $l_i$ and $l_{s}$ 
and the position of $l_i$ are computed as follows:}
\rv{
\begin{eqnarray}
\alpha(l_{i}) = 1 - e(t_{current})
\label{eq:superal}
\end{eqnarray}
\begin{eqnarray}
\alpha(l_{s}) = e(t_{current})
\label{eq:superas}
\end{eqnarray}
\begin{eqnarray}
\vec{p}(l_{i}) = \vec{p}_{start}(l_i) + (\vec{p}(l_{s}) - \vec{p}_{start}(l_i)) * e(t_{current})
\label{eq:superpos}
\end{eqnarray} }

%% file: sections/result.tex
\section{Experimental Results}
\label{sec:result}

To assess the applicability of our technique, we investigate three different use cases, including a (1) \emph{Synthetic Dataset}, a (2) \emph{Local Shops Dataset}, and the (3) \emph{Tokyo Disneyland Dataset}. 
The \emph{Synthetic Dataset} shows different variations of label layouts.
The \emph{Local Shops Dataset} provides a real-world example, where the labels are close and next to each other. 
The \emph{Tokyo Disneyland Dataset} presents another real-world scenario, \rv{where the labels are spread out in the 3D scene.} 
We use Unity as the visualization platform~\cite{unity} and incorporate the Vuforia engine~\cite{vuforia} to arrange objects in AR. The images shown in this section were taken using a Xiaomi Mi A2 device (Qualcomm Snapdragon $660$ processor and $4$ GB RAM) \rv{with} Android $10$ in portrait mode.

\subsection{Synthetic Dataset}
\label{subsec:syntheticdata}

\begin{figure}[tb!]
\centering
\setlength{\tabcolsep}{1pt}
\begin{tabular}{ccc}
\includegraphics[width=0.33\linewidth]{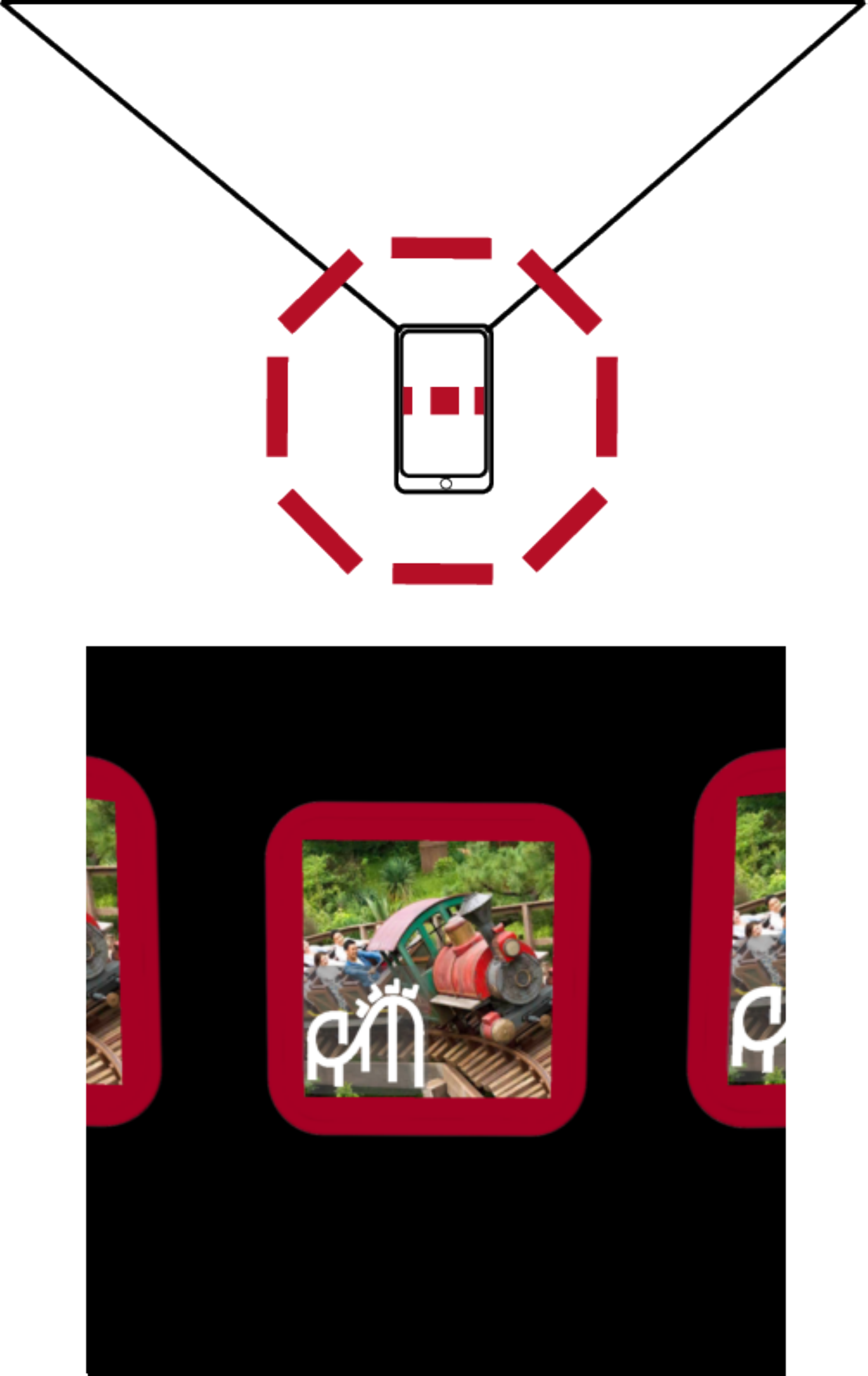} &
\includegraphics[width=0.33\linewidth]{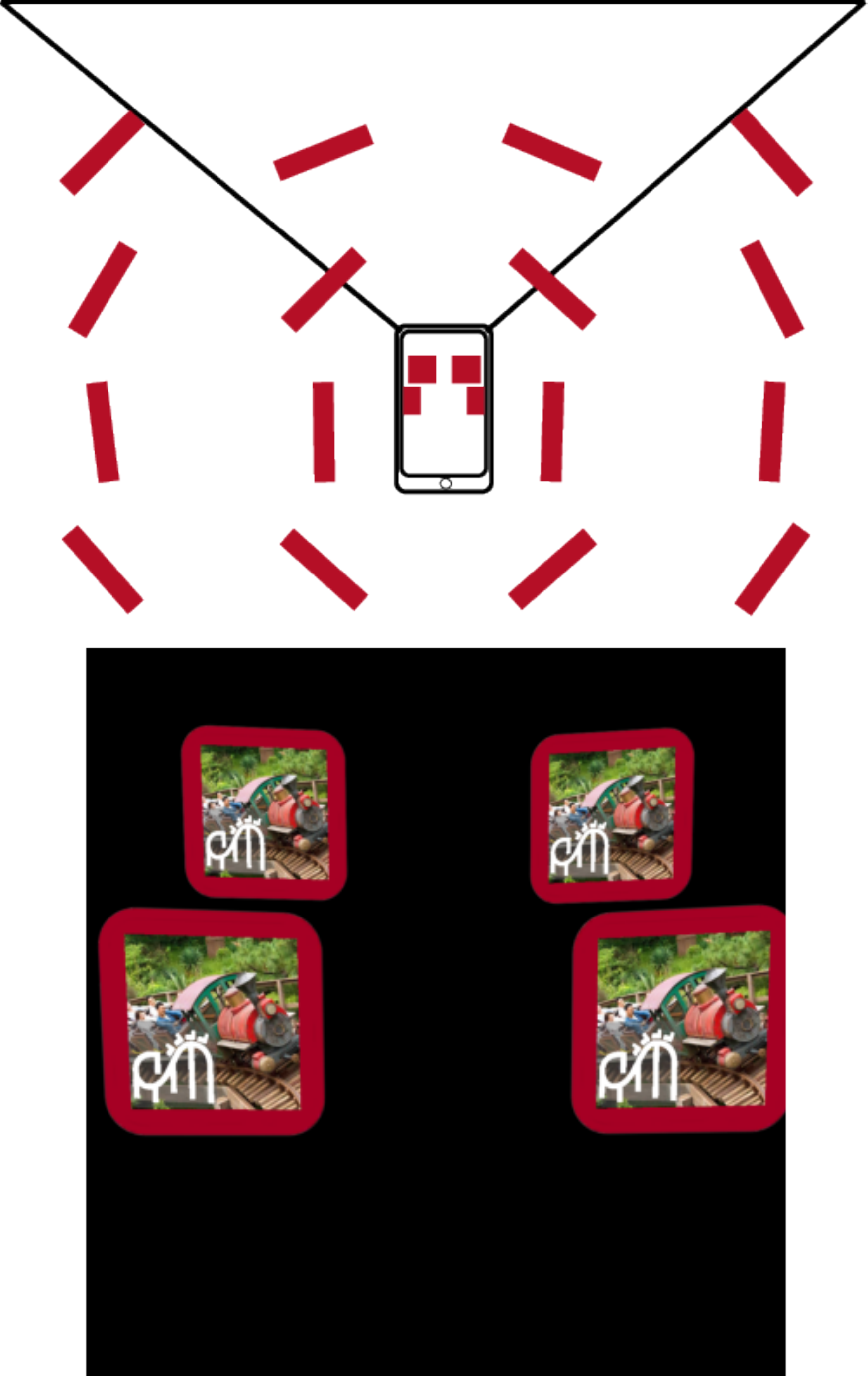} &
\includegraphics[width=0.33\linewidth]{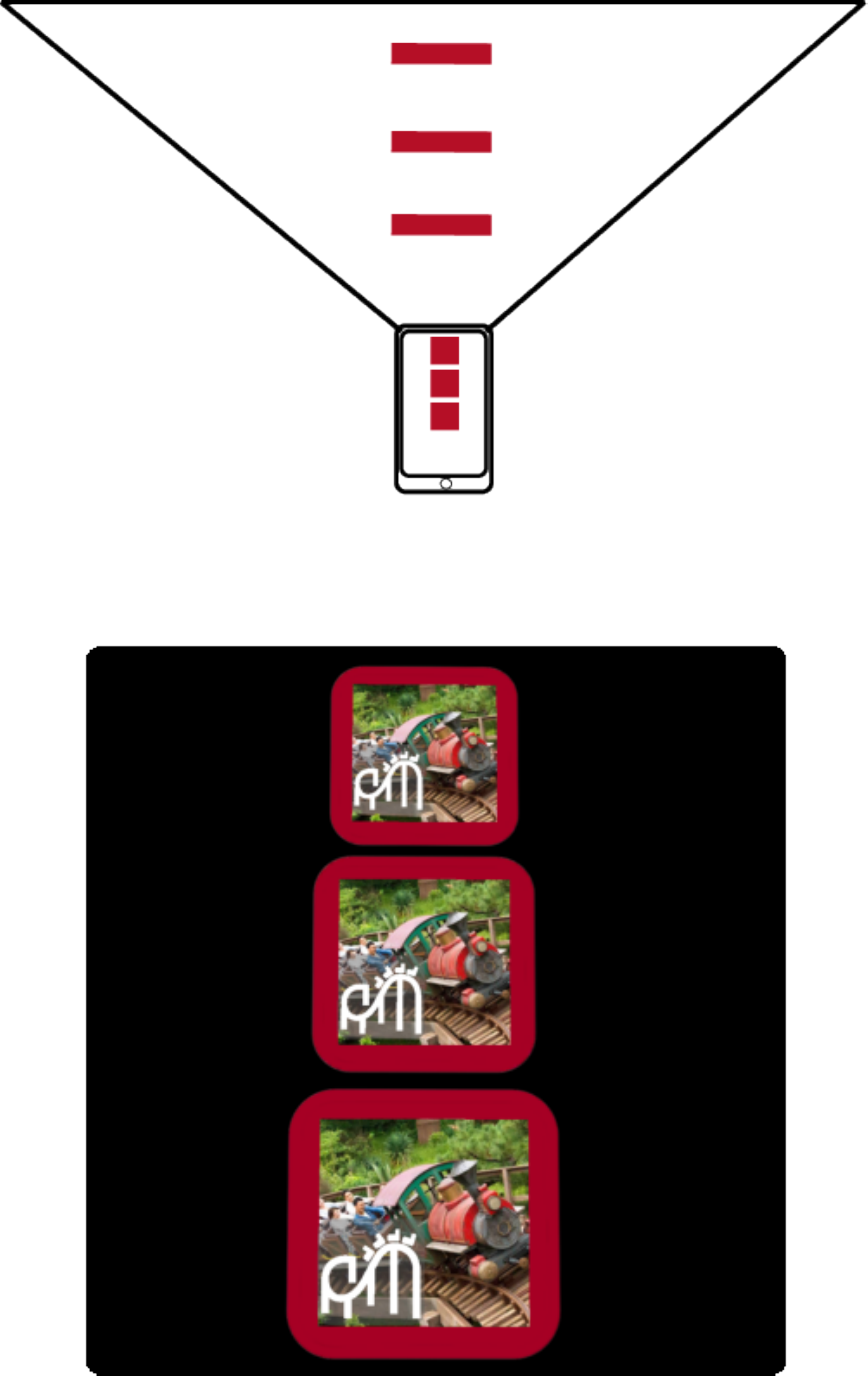} \\
 (a) & (b) & (c) \\
\end{tabular}
\caption{An example of the \emph{Synthetic Dataset} in top view with the displayed results beneath. Labels are arranged on a \rv{(a) circle, (b) grid, and (c)line.}}
\label{fig:syntheticlayout}
\vspace{2mm}
\centering
  \includegraphics[width=\linewidth]{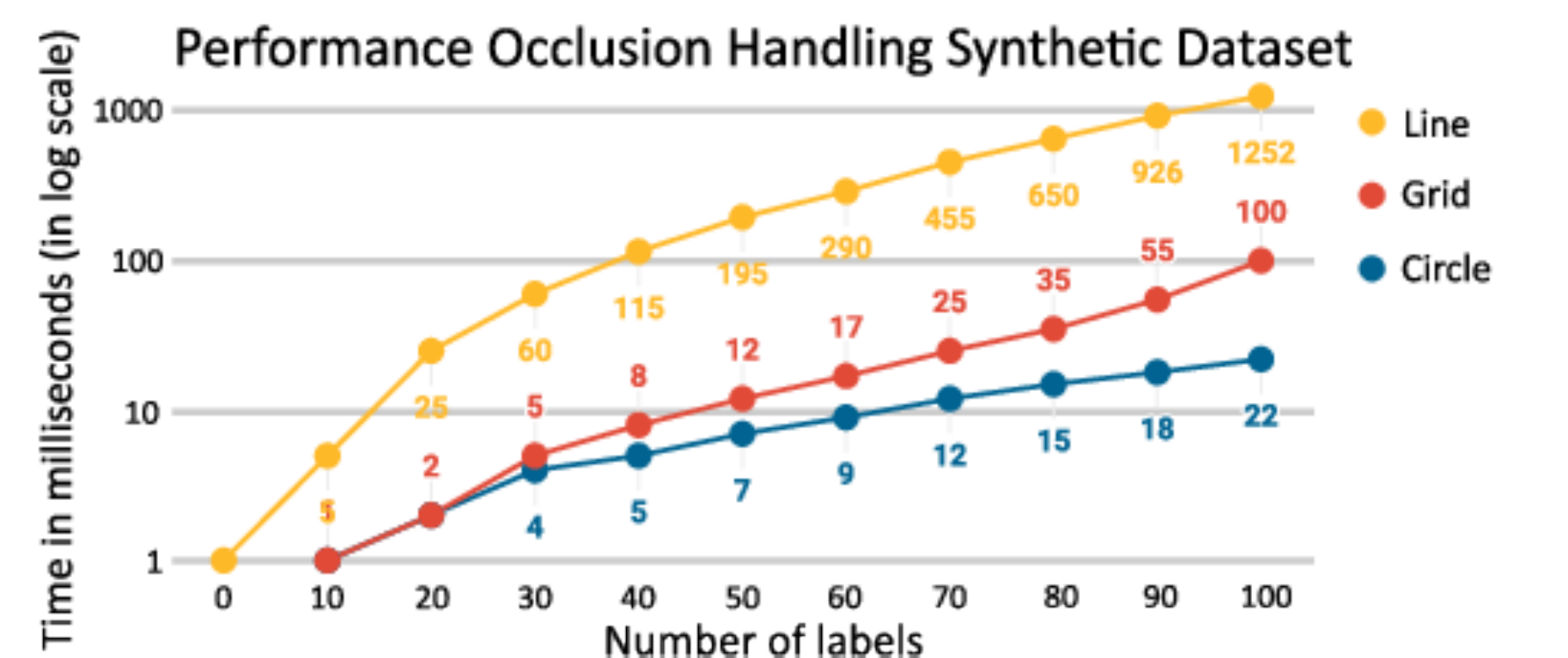}
  \caption{\rv{Computation times for removing occlusions}.}
  \label{tab:computationtimes}
\end{figure}

We \rv{study three different label layouts of} the \emph{Synthetic Dataset} (Figure \ref{fig:syntheticlayout}) and compute the execution time measured on the mobile device Xiaomi Mi A2. 
The three layouts are a circle layout, a grid layout, and a line layout, which are \rv{computationally increasingly expensive.} 
This assumption is based on the fact that if more labels are hidden in the current viewing direction, more occlusion removal steps are necessary.
Figure~\ref{tab:computationtimes} gives the execution times of all layouts in milliseconds based on a variation of label numbers. 
The labels in this dataset have a height and width of $120$ world space units \rv{by default in Unity.}

The circle layout (Figure~\ref{fig:syntheticlayout}(a)) requires the least computation times to resolve occlusions since many labels are initially \rv{arranged without occlusion issues}. The radius of the circle layout is set to $1,000$ world space units \rv{in this experiment.}
The grid layout (Figure~\ref{fig:syntheticlayout}(b)) distributes the labels equally leading to densely \rv{placed labels} in the scene. 
\rv{In our setting, the} number of labels per row is equal to $\sqrt{n}$, where $n$ is \rv{the total number of labels in} Figure~\ref{tab:computationtimes}. 
If $\sqrt{n}$ is not an \rv{integer}, the layout contains one partial label row in the grid. The size of the grid is \rv{$4,000 \times 4,000$} world space units and includes both near and far labels in \rv{the} world space.
The line layout (Figure~\ref{fig:syntheticlayout}(c)) represents the worst case \rv{example.}
The labels are located \rv{one after another, which leads to the maximum number of $i-1$ shifts} for each label $l_i$. The labels are placed $90$ world space units behind each other.
As shown in Figure \ref{tab:computationtimes}, resolving occlusions for the grid layout leads to higher computation times than the circle layout, \rv{but} lower computation times compared to the line layout.

\begin{comment}

\begin{table*}[tbh!]
% \footnotesize
\scriptsize
% \tiny
\centering
\begin{tabular}{|c|c|c|c|c|c|c|c|c|} 
\hline
\multicolumn{9}{|c|}{\textbf{\emph{Synthetic Dataset} Occlusion Handling Performance}} \\
\hline
\hline
\multicolumn{3}{|c|}{\textbf{Circle Layout}} &  \multicolumn{3}{|c|}{\textbf{Grid Layout}} & \multicolumn{3}{|c|}{\textbf{Line Layout}}\\
\hline
\hline
\textbf{\#Labels} & \textbf{Time (ms)} & \textbf{\#Raycasts} & \textbf{\#Labels} & \textbf{Time (ms)} & \textbf{\#Raycasts} & \textbf{\#Labels} & \textbf{Time (ms)} & \textbf{\#Raycasts} \\ 
\hline
\hline
10 & 1 & 36 & 10 & 1 & 40 & 10 & 5 & 180 \\ 
\hline
20 & 2 & 76 & 20 & 2 & 80 & 20 & 25 & 760 \\ 
\hline
30 & 4 & 116 & 30 & 5 & 160 & 30 & 60 & 1696\\ 
\hline
40 & 5 & 156 & 40 & 8 & 244 & 40 & 115 & 3076 \\ 
\hline
50 & 7 & 196 & 50 & 12 & 324 & 50 & 195 & 4860 \\ 
\hline
75 & 12 & 296 & 75 & 43 & 804 & 75 & 570 & 11060 \\ 
\hline
100 & 20 & 396 & 100 & 124 & 2012 & 100 & 1340 & 19760 \\ 
\hline
\end{tabular}
\caption{Computation times for the occlusion handling measured on a Xiaomi Mi A2 smartphone for different synthetic label layouts.}
\label{tab:computationtimes}
\end{table*}
\end{comment}

\subsection{Local Shops Dataset}
\label{subsec:shopsdata}

The \emph{Local Shops Dataset} contains shop locations, types of shops, and number of people inside a shop (per m$^2$) of a strip mall (Figure~\ref{fig:shop}).
The icons indicate the respective shop types (e.g., clothing, shoes, and groceries). 
Considering the current COVID-19 regulations, we encode the number of people per m$^2$, to identify the customer density \rv{or COVID-19 safety measure} in the shop in real-time. 
In Figure~\ref{fig:shop}, we use a color scale from white to red. 
\rv{The text displays the name and measure accordingly.
Figure~\ref{fig:shop} gives} an explanatory result, in which the device is tilted. 
\rv{As shown here, the placement of the labels is thereby not influenced.}
The rectangular labels remain parallel to the ground. 

\subsection{Tokyo Disneyland Dataset}
\label{subsec:disneydata}

The \emph{Tokyo Disneyland} is one of the most popular amusement parks in the world. 
\rv{Many visitors often need to line up for hours to enjoy a specific attraction,} and many magazines and blogs guide visitors to optimize their one-day visit~\cite{Bricker:2020:dtb}.
The amusement park consists of $35$ big attractions, which we mark all as POIs \rv{in our system} to give an overview of the park.
In the park, themed areas, such as the \emph{Westernland}, are \rv{subregions grouping several attractions for convenience.} We use the themed areas of the amusement park to aggregate labels \rv{and present the area using the corresponding super label.}

\begin{figure}[tb!]
% \centering
% \setlength{\tabcolsep}{1pt}
% \renewcommand*{\arraystretch}{0.7}
\setlength\arraycolsep{0pt}
  \begin{minipage}[b]{0.18\textwidth}
    \centering{
      \includegraphics[width=\linewidth]{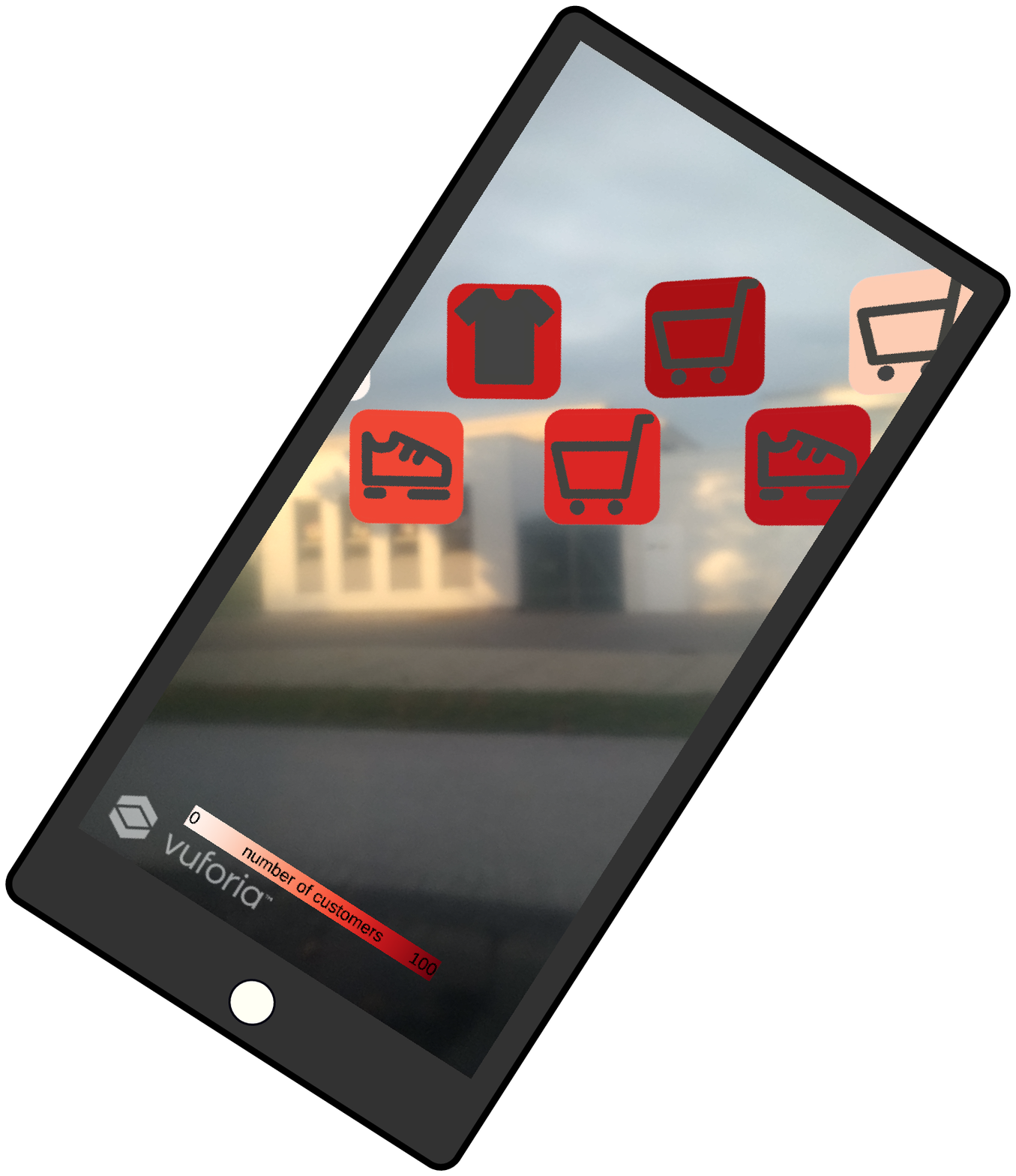} \\
    }
    \vspace{5mm}
    \caption{\rv{An example with a $45^{\circ}$~tilted mobile device}.
    \vspace{0.5mm}
    }
  \label{fig:shop}
  \end{minipage} \quad
  \begin{minipage}[b]{0.275\textwidth}
    \centering{
        \begin{tabular}{cc}
        \setlength{\tabcolsep}{0pt}
        \includegraphics[width=0.43\linewidth]{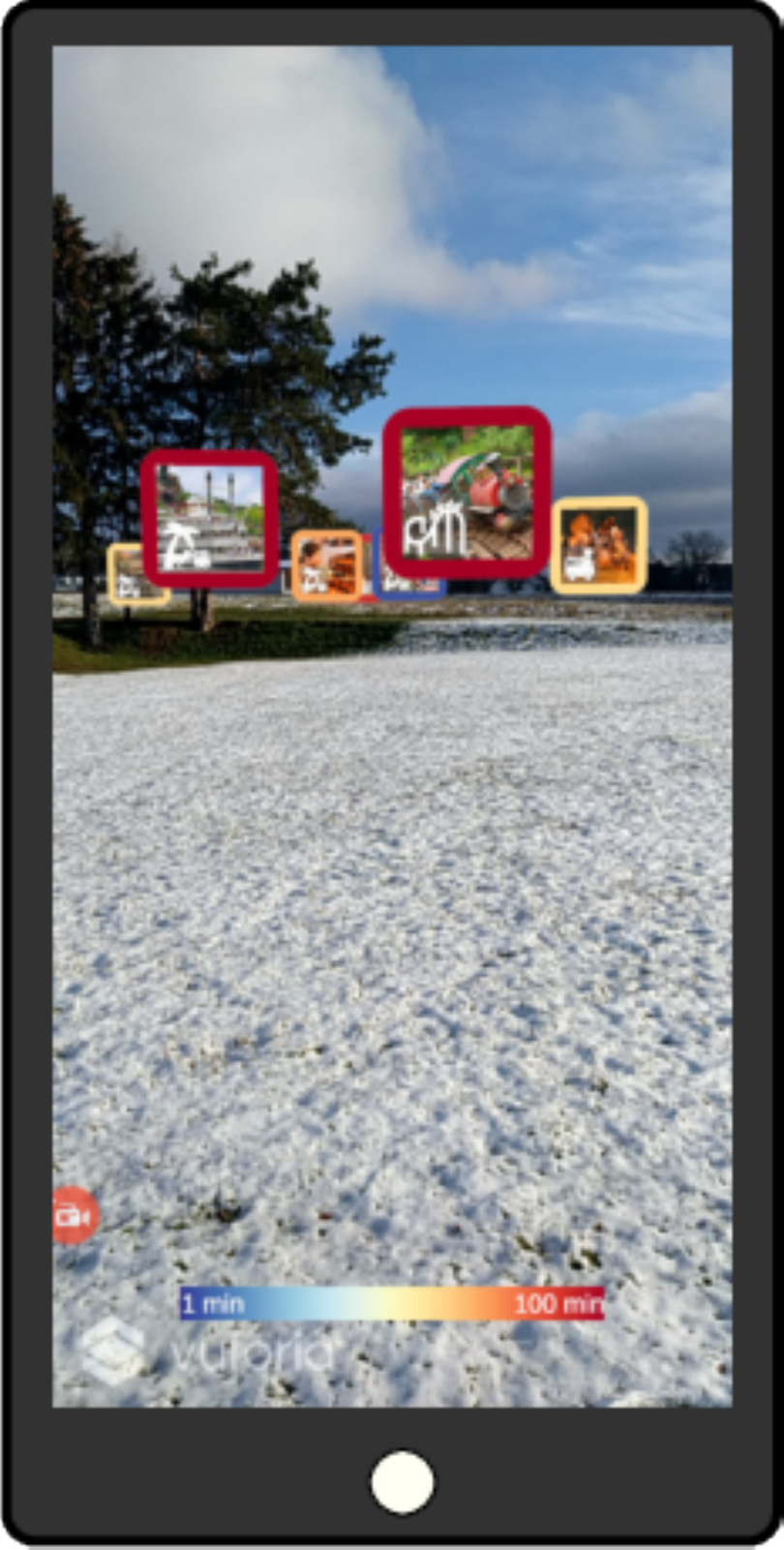} &
        \includegraphics[width=0.43\linewidth]{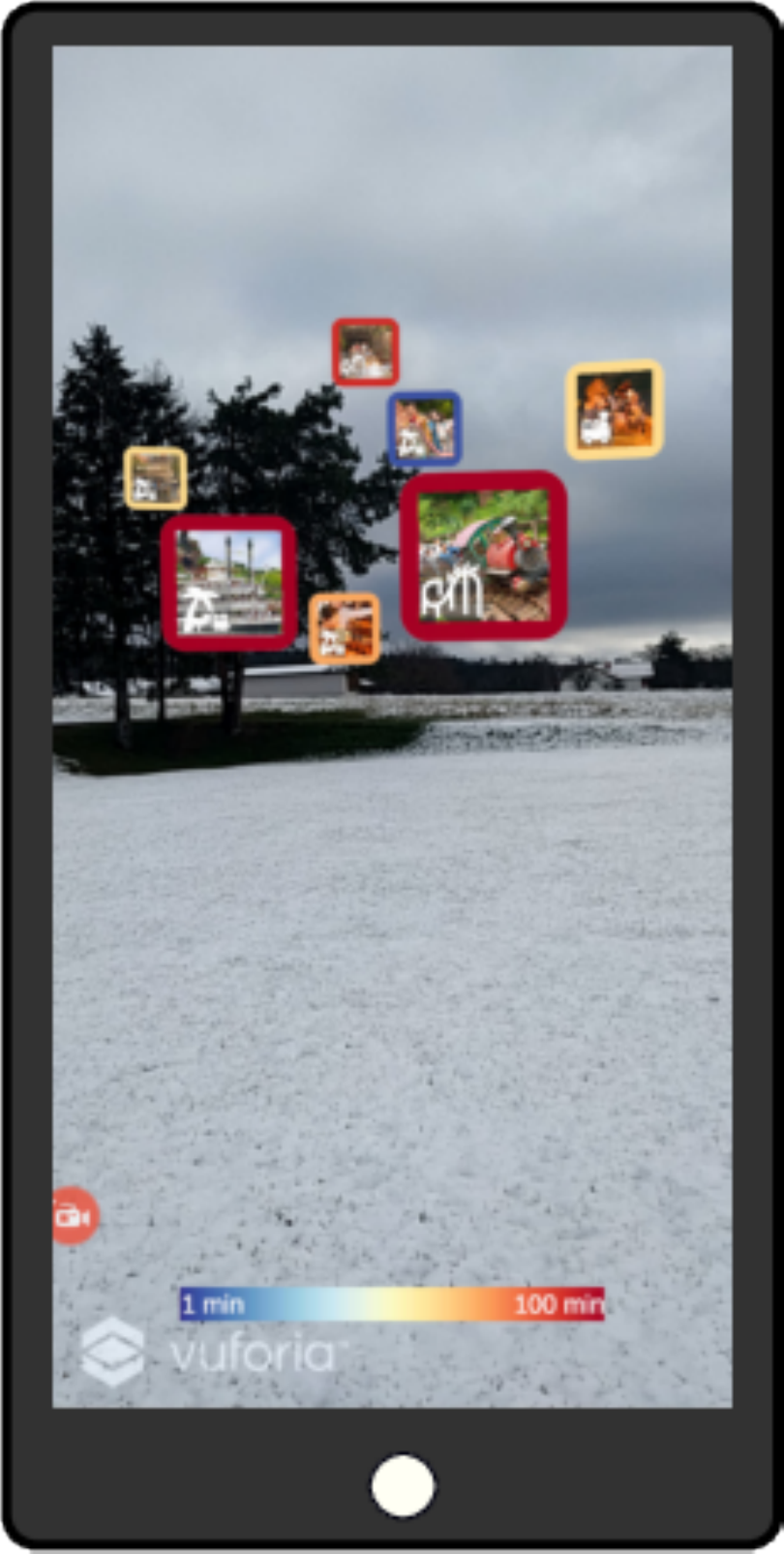} \\
         \rv{(a)} & \rv{(b)} \\
        \end{tabular}
    }
    \caption{Occlusions that occur in (a) are resolved in (b).}
  \label{fig:comp12}
  \end{minipage}%

\end{figure}

\begin{figure*}[th!]
\centering
  \includegraphics[width=0.95\linewidth]{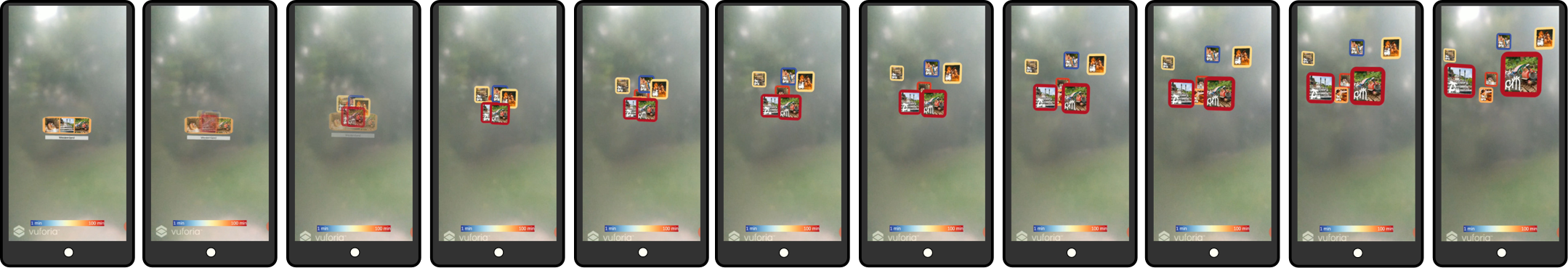}
  \caption{Transition from a super label to the individual labels for each POI over time.}
  \label{fig:areatransition1}
\end{figure*}

\rv{Once the \emph{positioning labels in AR} has been preprocessed, labels might} initially be occluded. 
Figure~\ref{fig:comp12} compares the results of the same position and viewing angle.
\rv{Initially, the labels are occluded as shown in Figure~\ref{fig:comp12}(a) and the respective occlusion-free result is \rv{given} in Figure~\ref{fig:comp12}(b).}
\rv{Since the occlusion-free results are independent of} the viewing angle of the device, no incoherent label movement occurs when the user rotates the device. The occlusions are resolved for all the labels around the users as explained in Section~\ref{sec:overview} and Section~\ref{ssec:occclusion}. 
Labels closer to the user are more likely to stay close to their initial positions than labels that are farther away. 
\rv{The two closest labels in Figure~\ref{fig:comp12}} are \emph{Big Thunder Mountain} and \emph{Mark Twain's Riverboat} showing an iconic image of a train and a boat. 
The positions of these two labels are not changed. 
\rv{Labels that are occluded by these two labels will be shifted upwards during the \emph{occlusion management}.}
Figure~\ref{fig:areatransition1} depicts \rv{the transition of a super label to its} individual labels. 
\rv{The super label represents the} \emph{Westernland} themed area of the \emph{Toko Disneyland}.

\begin{figure*}[tbh!]
\begin{minipage}{0.80\textwidth}
\centering{
 \setlength{\tabcolsep}{1pt}
 \begin{tabular}{cccc}
   %%%%%%%%%%
   %\includegraphics[width=0.18\linewidth]{images/results/LocalShops.eps}
    \includegraphics[width=0.24\linewidth]{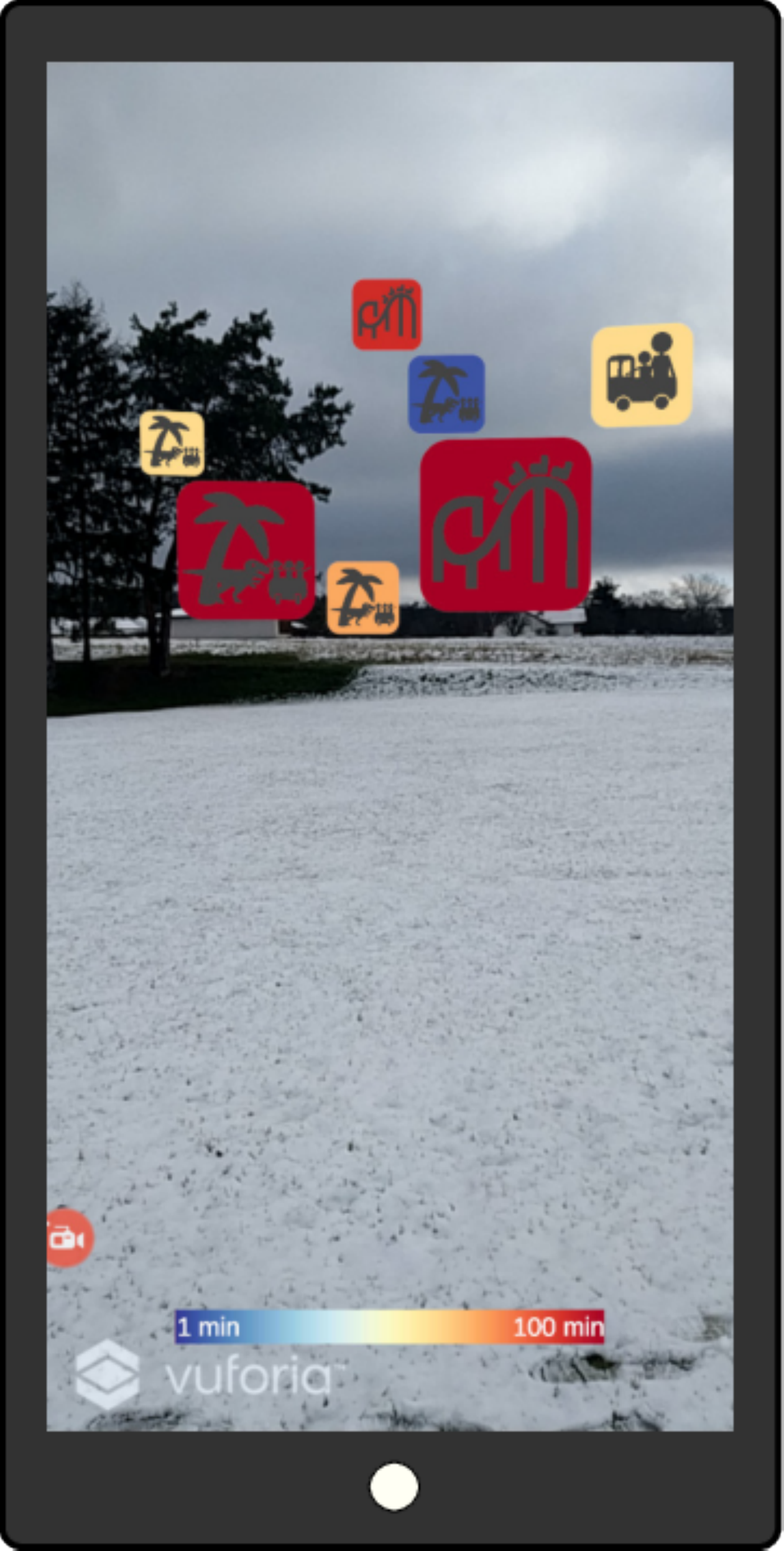} &
    \includegraphics[width=0.24\linewidth]{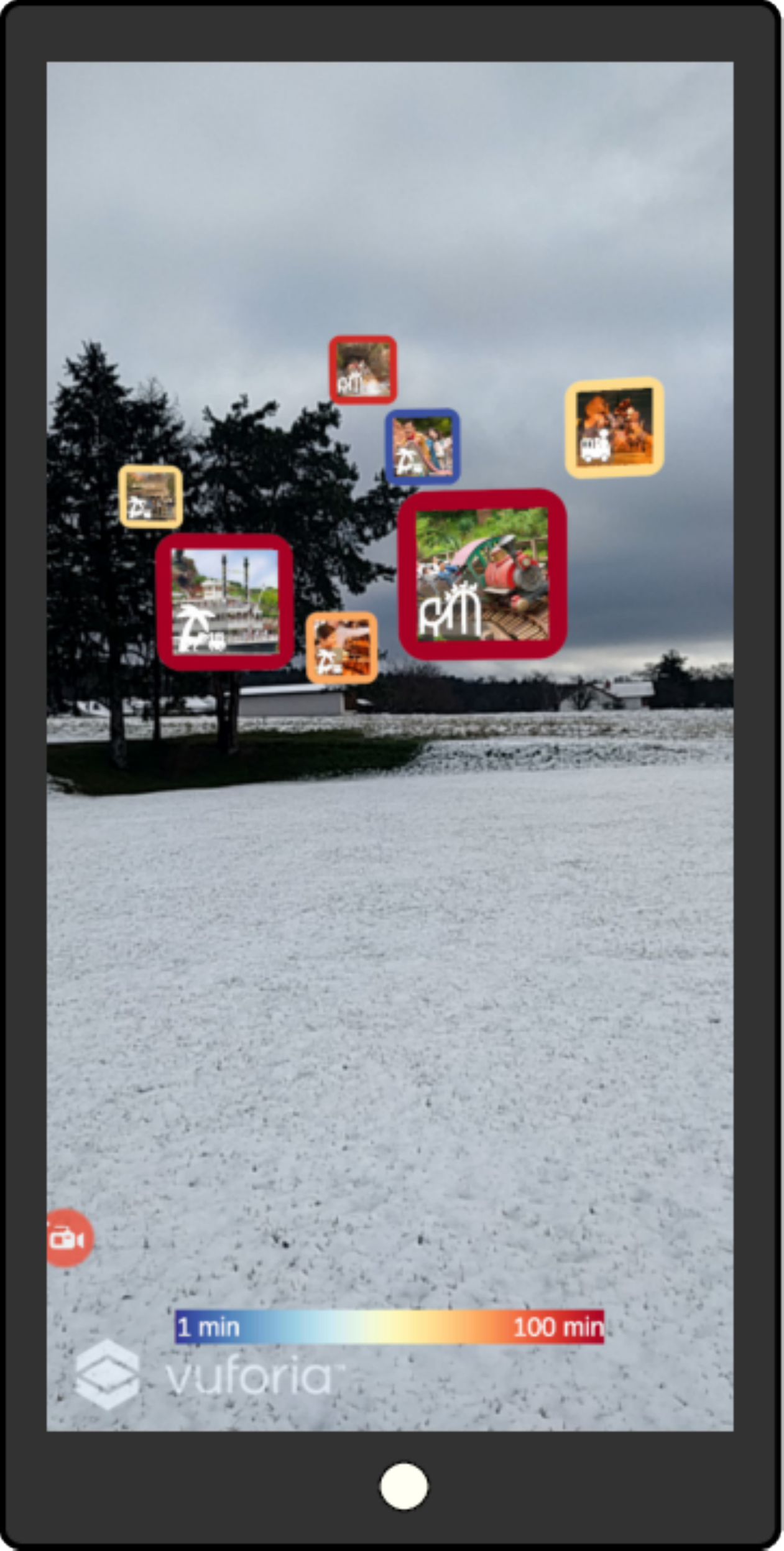} &
    \includegraphics[width=0.24\linewidth]{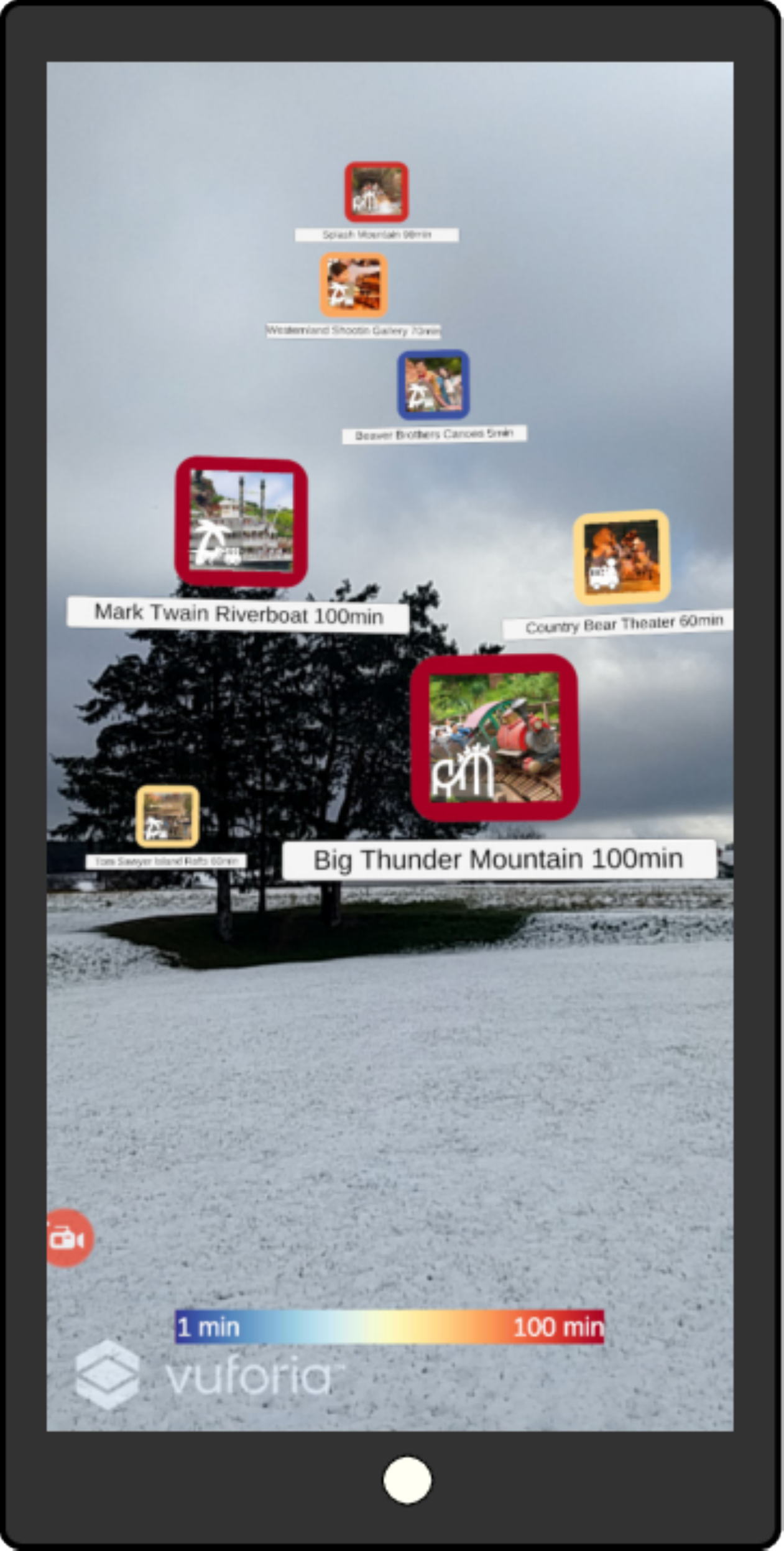} &
    \includegraphics[width=0.24\linewidth]{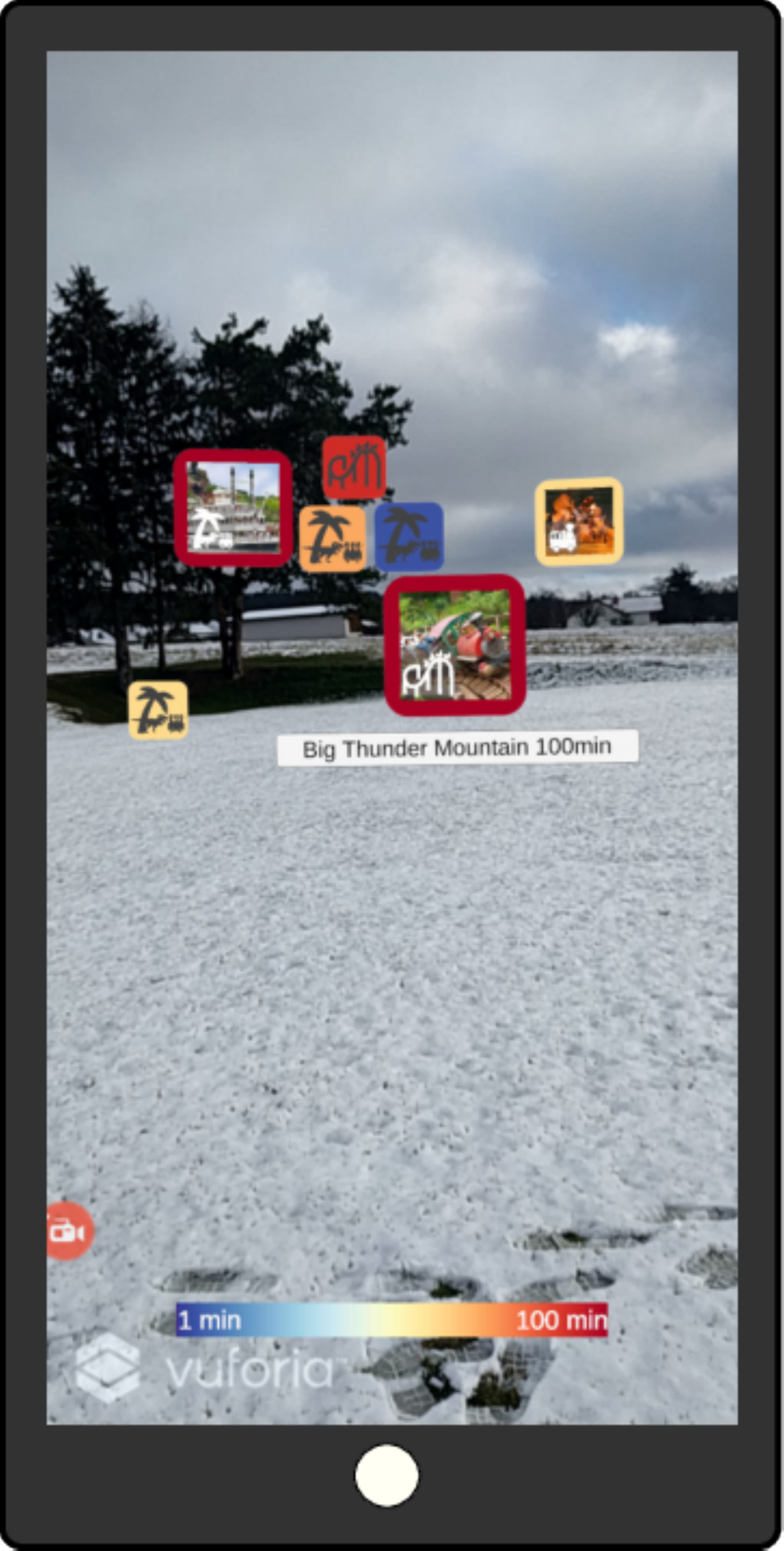} \\
    % \includegraphics[width=0.25\linewidth]{images/results/Shops/RotationDifference.eps} \\
   %%%%%%%%%%
   \rv{(a) Lowest LOD} & 
   \rv{(b) Middle LOD} & 
   \rv{(c) Highest LOD} & 
   \rv{(d) Dynamic LODs} \\
%   (e) Tilted device \protect\emph{Local Shops Dataset}\\
 \end{tabular}
}
\caption{A comparison of different LODs and dynamic LODs (\rv{applying} the \protect\emph{level-of-detail management})}
% , and an explanatory result with a tilted device side by side.}
 \label{fig:compareLODsTilt}
\end{minipage}\hfill % maximize space 
\begin{minipage}{0.2\textwidth}
\centering{
 \setlength{\tabcolsep}{1pt}
 \begin{tabular}{cc}
   %%%%%%%%%%
   %\includegraphics[width=0.18\linewidth]{images/results/LocalShops.eps}
    \includegraphics[width=0.48\linewidth]{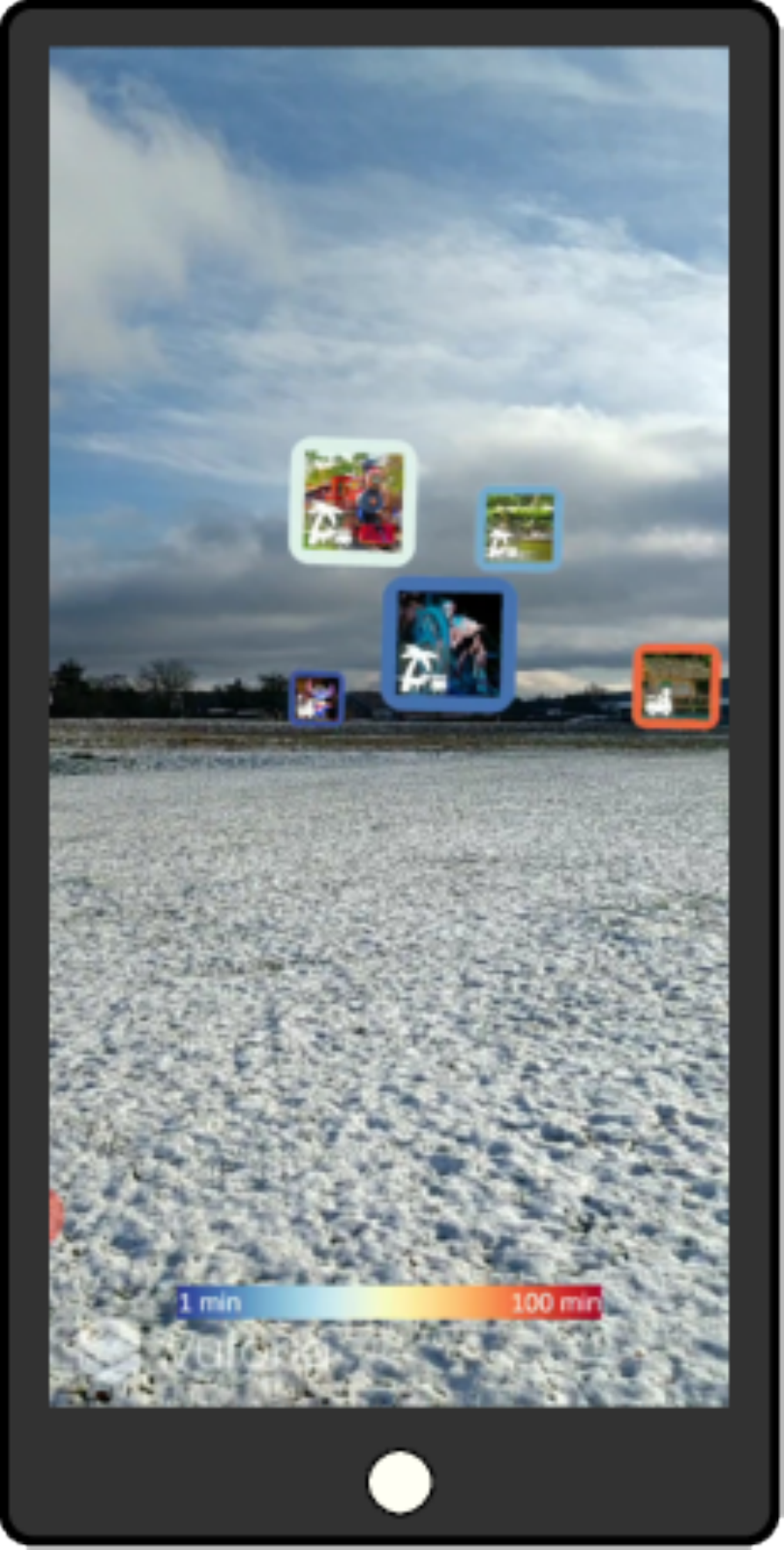} &
    \includegraphics[width=0.48\linewidth]{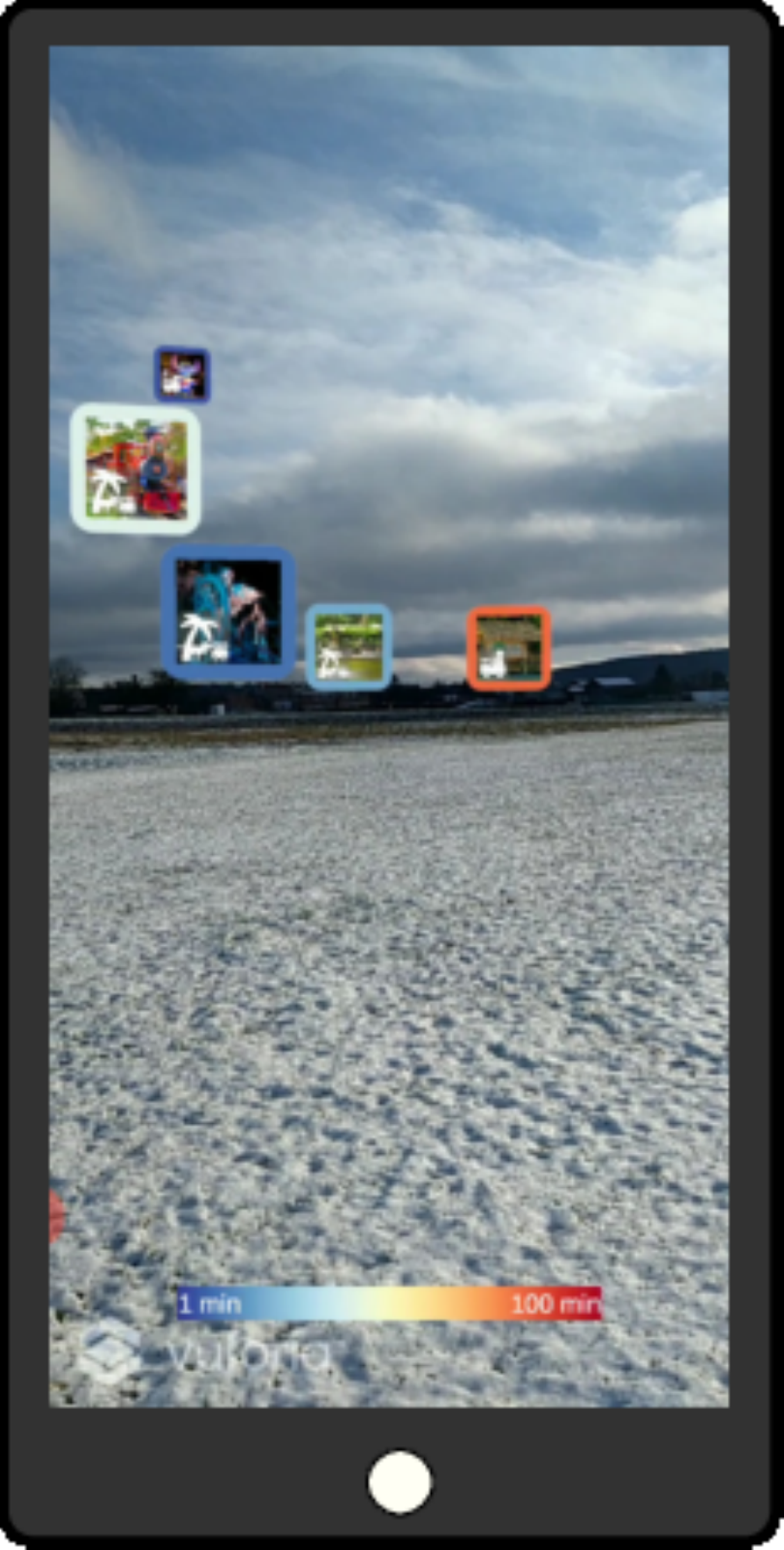} \\
    \rv{(a)} & \rv{(b)}\\
    \includegraphics[width=0.48\linewidth]{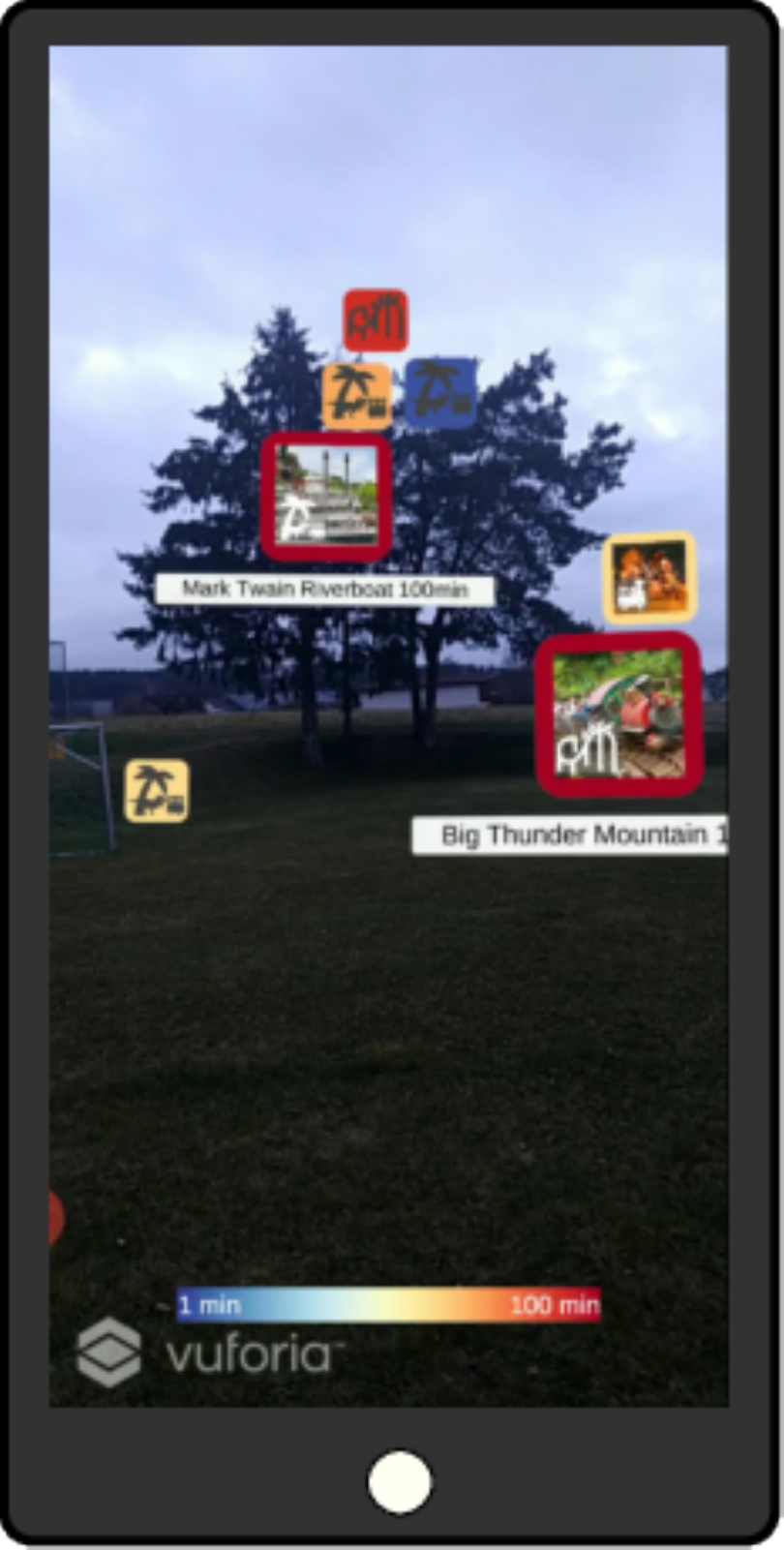} &
    \includegraphics[width=0.48\linewidth]{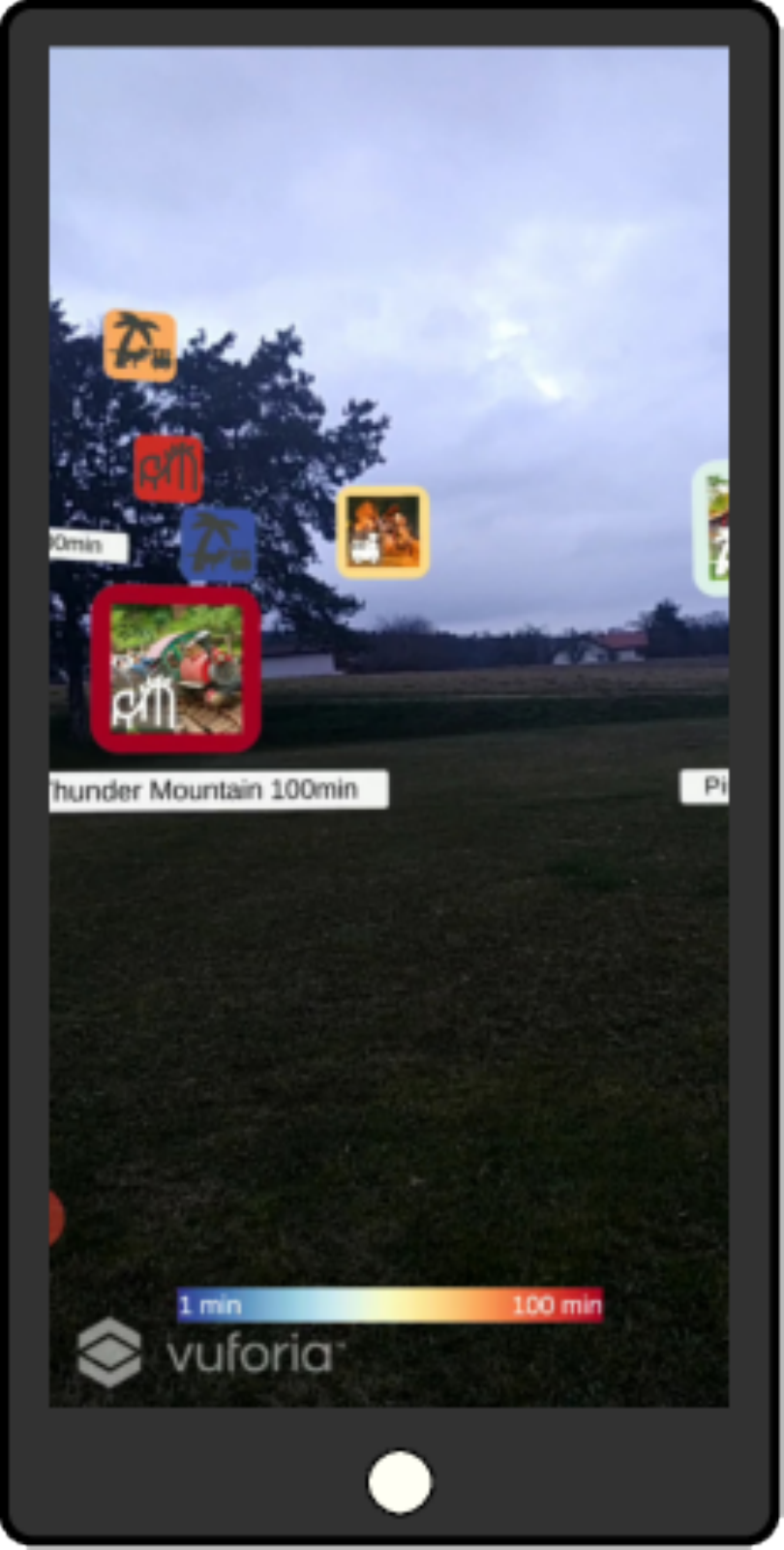} \\
    \rv{(c)} & \rv{(d)} \\
    \vspace{-7mm}
   %%%%%%%%%%
   %\rv{(a) Lowest LOD} & 
   %\rv{(b) Middle LOD} & 
   %\rv{(c) Highest LOD} & 
   %\rv{(d) Dynamic LODs} \\
%   (e) Tilted device \protect\emph{Local Shops Dataset}\\
 \end{tabular}
}
\rv{
\caption{Lateral transitions}
\vspace{2mm}
\label{fig:latTransNew}
}
% , and an explanatory result with a tilted device side by side.}

\end{minipage}
\end{figure*}

Figure~\ref{fig:compareLODsTilt} \rv{presents different} LODs of the themed area \emph{Westernland}. 
Figure \ref{fig:compareLODsTilt}(a) \rv{shows} all labels in the \rv{lowest} LOD consisting of a colored rectangle encoding the waiting time and an icon indicating the attraction type. 
This LOD provides the \rv{simplest} overview of the attractions, \rv{and} it presents the least amount of information as only the attraction \rv{types} and the color encodings are included. Figure \ref{fig:compareLODsTilt}(b) illustrates the middle LOD adding an iconic image to the encoding. 
In this case, the type icon is less dominant than in the \rv{lowest} LOD. 
Figure \ref{fig:compareLODsTilt}(c) depicts the \rv{highest} LOD by adding a text tag stating the name and the exact waiting time of the attraction in minutes. 
This LOD provides the most detailed information. However, higher vertical stacking of labels is necessary to resolve occlusions compared to the \rv{lowest} and middle LOD during the occlusion handling. 
Figure \ref{fig:compareLODsTilt}(d) \rv{presents the label placement of the themed area \emph{Westernland} once the dynamic LOD selection is enabled.}
This solution constitutes a compromise \rv{concerning} the presented amount of information \rv{and label displacement. It includes} detailed information about close attractions and \rv{keeps} the vertical stacking of labels low compared to the \rv{highest} LOD. 
The preferred LOD might vary depending on the use case and the user's preference (see Section~\ref{sec:evaluate}). 
Each LOD has its benefits and drawbacks with the dynamic LODs being the most versatile one as they present detailed information about close labels and avoid excessive vertical stacking (see Section~\ref{sec:evaluate}). \rv{Figure~\ref{fig:latTransNew} exemplifies lateral translations of the user and the resulting label arrangements. Figure~\ref{fig:latTransNew}(a) and Figure~\ref{fig:latTransNew}(c) correspond to the initial positions. In Figure~\ref{fig:latTransNew}(b) and Figure~\ref{fig:latTransNew}(d), the user moved laterally to the right. The label positions are updated smoothly depending on the movement of the user.}

%% file: sections/evaluate.tex
\section{Qualitative Evaluation}
\label{sec:evaluate}

We conducted an online survey to evaluate the effectiveness \rv{and the applicability of our approach. 
Primarily, we aim to confirm the appropriateness of the selected design principles. It is based on users' preferences by examining task performance in terms of \rv{required} time and result accuracy.
Our hypotheses of the study are summarized as follows:
\begin{itemize}
    \item[\textbf{(H1)}] The design principle, removing label occlusions, has higher priority in comparison to showing precise positions of labels.
    \item[\textbf{(H2)}] Rich label design in AR leads to a better POI exploration and decision-making experience in contrast to plain text labels.
    % \item[\textbf{(H3)}] Aggregated super labels provide a visual summary of regions containing several POIs.
    \item[\textbf{(H3)}] Users can perform faster route planning tasks using our system \rv{compared to} conventional maps.
\end{itemize}
}

\rv{We further decompose our hypotheses into four main tasks as summarized in Table~\ref{tab:tasks} \rv{for} an online questionnaire.
In the future, we plan to do an in-person user study as one of our primary attempts.
% The user study consists of four main tasks of which each question requires participant input. 
For each measurable task, time and accuracy were collected. 
After each task, we also asked participants to provide reasons regarding their experience when performing the task.
At the end of the entire questionnaire, we requested general feedback and collected some personal information for further analysis (e.g., age, educational background, experience with AR devices, and so forth).
Privacy agreements have been received prior to the user study and the collected data is carefully stored without identifications of the participants.
In total, we recruited \rv{$30$} participants who are experienced with visualization techniques and graduate students of visual computing participated in the survey. 
The age of the participants ranges from 24 to 64 years with the majority of participants being in the late twenties or the early thirties.
One limitation of the user study comes from the limited access to the general audience, while experience in visual computing will help the participants to answer the questions smoothly. 
We performed a within-subjects study design, where we \rv{tested} all variable conditions for a participant in order to analyze individual behaviors in more depth.
Questions in each task are also randomized to avoid a learning effect. 
For more details, \rv{we refer} to the accompanying supplementary materials.
}

\begin{table}[ht!]
\centering
\scriptsize
\begin{tabular}{|c|l|} 
\hline
Tasks   & \rv{Goal of the investigation and question samples} \\ 
\hline
\hline
Task 1  & \rv{Impact of occlusion on attribute tasks and comparative tasks} \\
        & \rv{Q1: What is the waiting time of an attraction?} \\
        & \rv{Q2: Which attraction has the minimal waiting time?} \\
\hline
Task 2  & \rv{Effectiveness} of levels-of-detail \\ % (e.g., text labels, colored labels, and super labels)\\ 
        & \rv{Q3: Which themed area has the minimal waiting} \\
        & \rv{\hspace{5mm}time? (with LOD variations)} \\
        & \rv{Q4: Which LOD do you prefer?} \\
\hline
Task 3  & Effectiveness of 2D maps and our AR encoding \\ 
        & \rv{Q5: Choose the attraction with the minimal} \\
        & \rv{\hspace{5mm} waiting time in the specified themed area} \\
        %& \rv{Q6: Which LOD do you prefer when solving the problem?} \\
        % \hline
% Task 4 & Effectiveness of text labels, colored labels, and super labels  \\ 
\hline
Task 4  & Combinatorial features in our system \\ 
        & \rv{Q6: Provide your feedback to different configuration settings} \\
\hline
\end{tabular}
\caption{Overview of the tasks in the user study.}
\label{tab:tasks}
\end{table}

\begin{figure}[th!]
\centering
\setlength{\tabcolsep}{2pt}
    \begin{tabular}{c|c}
    \includegraphics[width=0.49\linewidth]{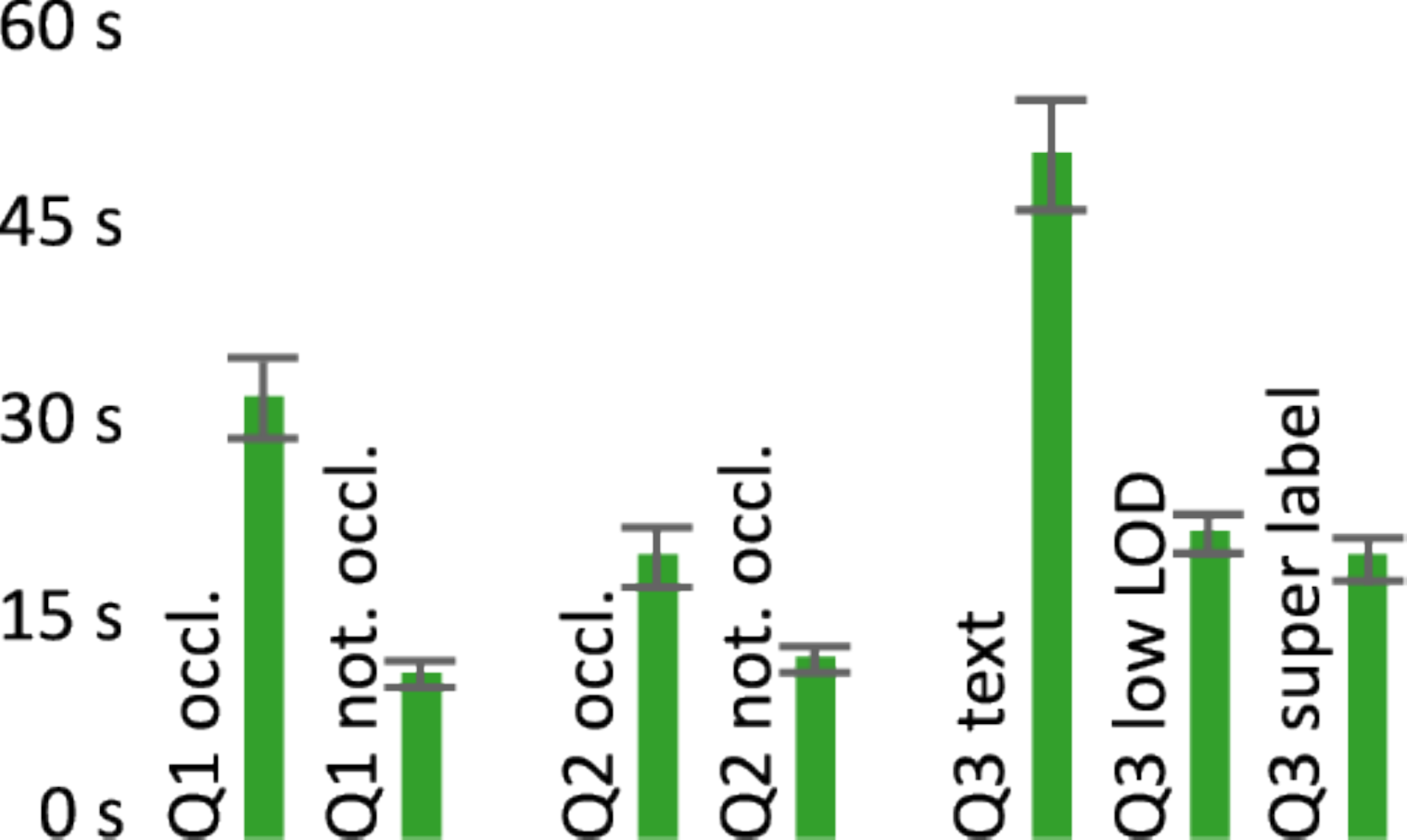} &
    \includegraphics[width=0.49\linewidth]{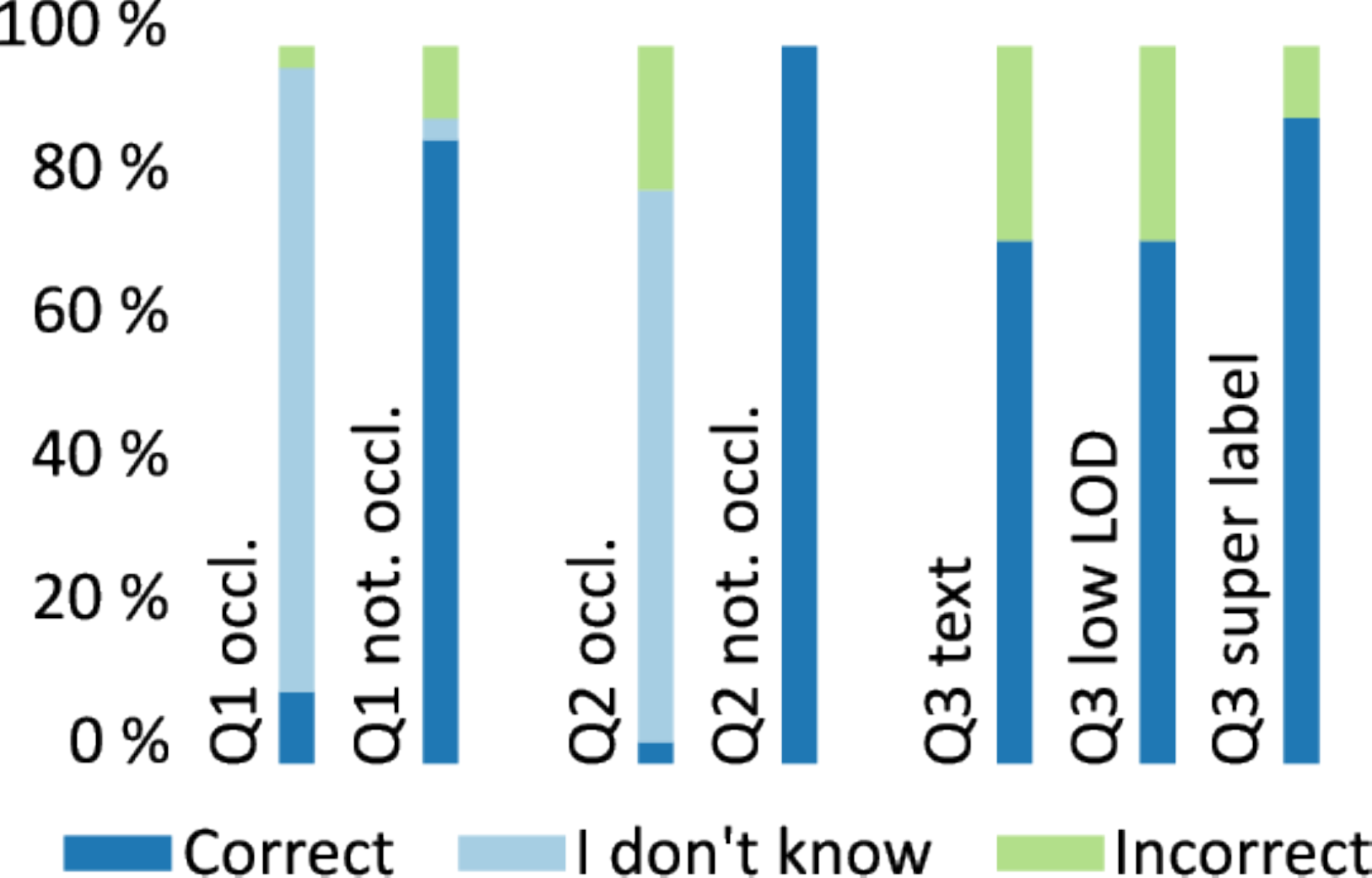} \\
     \rv{(a) Time} & \rv{(b) Accuracy} \\
    \end{tabular}
    \rv{\caption{(a) Task completion times (in seconds) and (b) accuracy of Q1 to Q3. The error bars represent the standard errors.}
    \label{fig:q1-3}
    }
\vspace{1mm}
    \centering{
        \begin{tabular}{c|c|c}
        \includegraphics[width=0.49\linewidth]{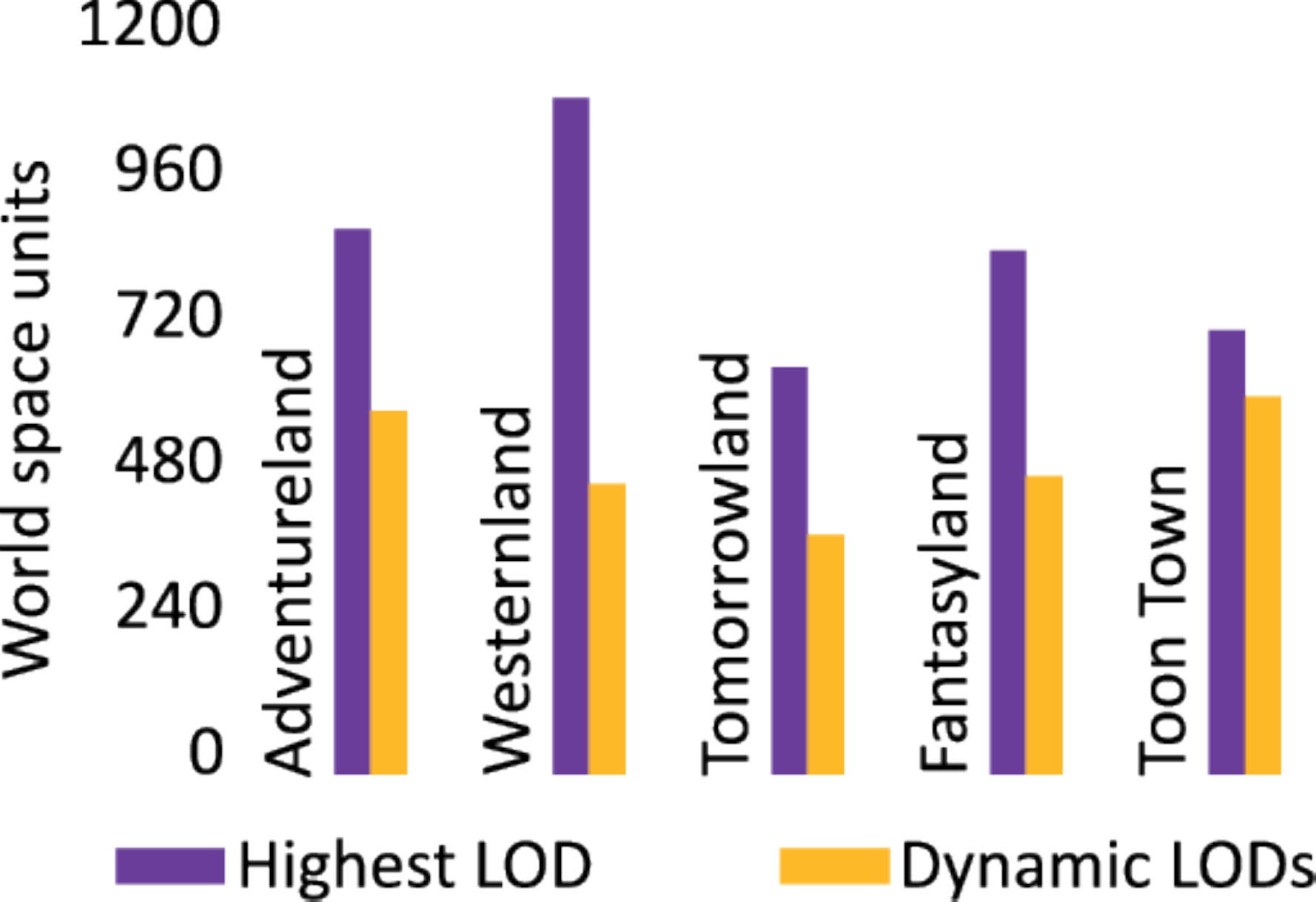}
         &
         \includegraphics[width=0.3\linewidth]{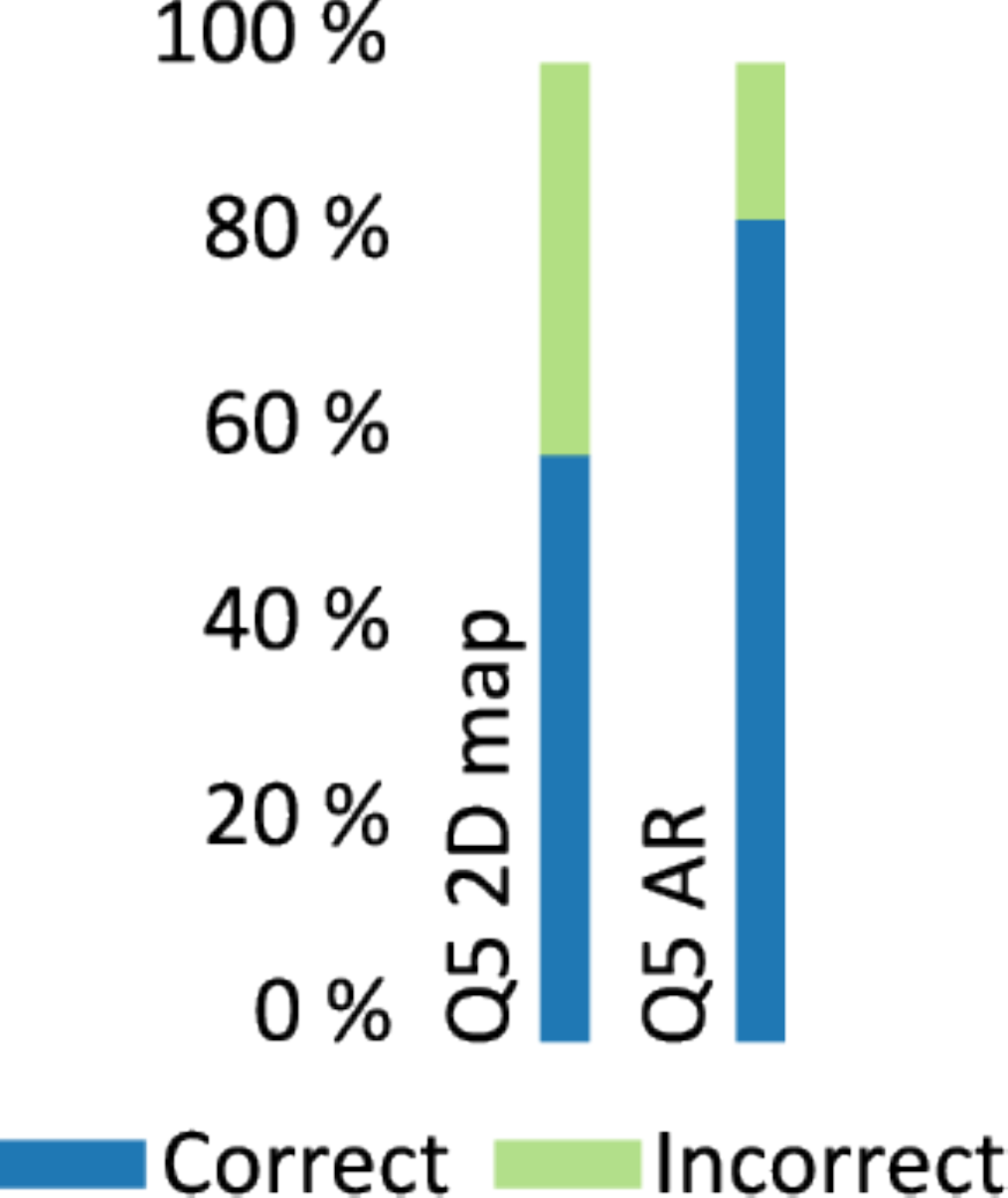}
         & 
         \includegraphics[width=0.16\linewidth]{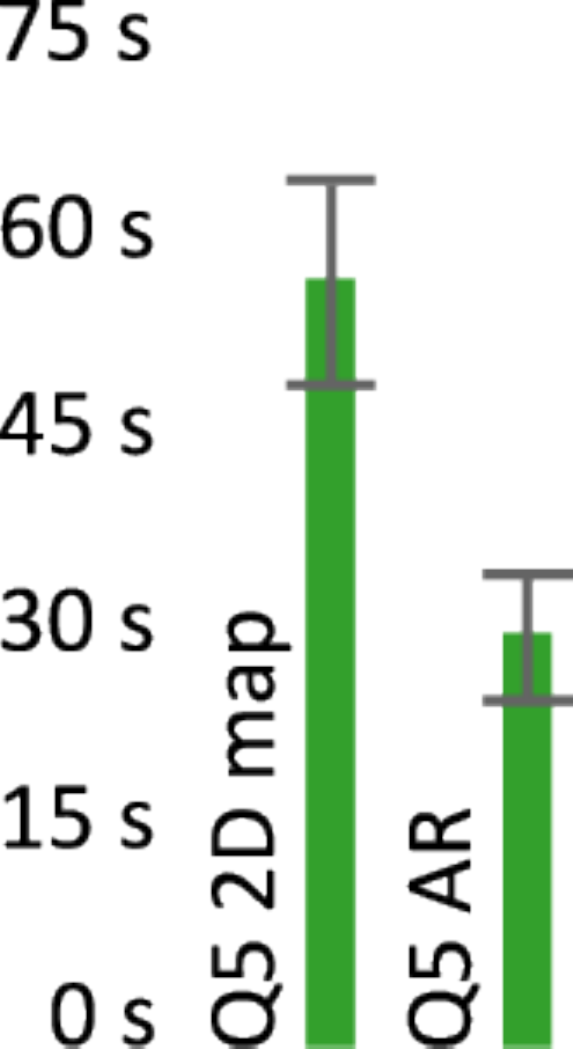}
         \\
        \rv{(a) Height of labels} & \rv{(b) Accuracy} & \rv{(c) Time} \\
        \end{tabular}
    }
    \rv{\caption{(a) Combined height of the stacked labels. (b) Accuracy and (c) task completion times (in seconds) of Q5. The error bars show the standard errors.}
    \label{fig:q5}
    }
    
\end{figure}

% The first hypothesis 
\textbf{(H1)} \rv{demonstrates the importance of resolving label occlusions in AR.} 
As described in Section~\ref{sec:related},
existing work concludes the importance of resolving occlusions in AR to support the decision making process by the users~\cite{grassetimage}. 
\rv{In Task~1, we show participants a few snapshots (see supplementary materials) of our system, and ask the participants to} determine the waiting time of the specified attraction (Q1) and select the attractions with minimal waiting times (Q2).
Three participants managed to select the correct waiting times if occlusions occurred, 
\rv{and the participants} stated that the waiting times were not recognizable \rv{in such a situation. 
Figure~\ref{fig:q1-3} \rv{summarizes} task completion time and accuracy. The time needed to answer the questions could be decreased (Q1 from $33.26~s$ to $12.6 ~s$, Q2 from $21.39~s$ to $13.7~s$) and the number of correct answers could be increased (Q1 from $10~\%$ to $86.67~\%$, Q2 from $3.33~\%$ to $100~\%$) when showing results with our \emph{occlusion management} (Figure~\ref{fig:q1-3}).} \rv{$24$} participants explicitly stated that it was difficult or impossible to select the correct answers \rv{if} information is occluded, and \rv{$24$} participants agree that the occlusion-free positioning eases decision-making processes when investigating the labels.

\rv{For hypothesis \textbf{(H2)}, we design questions in Task~2, where participants need to take several attributes} into account to answer the questions. 
\rv{In Q3, the participants were} asked to select a themed area of the amusement park with the lowest average waiting time. 
\rv{We showed participants images with labels of different LOD settings, 
including text labels, labels with the lowest LOD, and super labels.} The time needed to answer questions \rv{for} this task is summarized in Figure~\ref{fig:q1-3}(a). The participants, in general, spent more time if only text labels are present \rv{($52.05~s$ on average) since they probably like} to calculate the correct number to answer the questions properly. 
If we present information using the lowest LOD, a shorter time \rv{($23.52~s$ on average)} was required in comparison to pure text labels. \rv{Using super labels achieved a similar performance, participants spent} \rv{$21.61~s$} to answer the questions. %
If the waiting time is depicted using text labels or labels in the lowest LOD, the themed area with the minimal average waiting times was correctly selected by \rv{$73.33~\%$} of the participants. \rv{$90~\%$} of the participants selected the correct answers if the super labels were shown (Figure~\ref{fig:q1-3}(b)).

In Q4, we ask participants which LOD they prefer. We presented text labels, labels in one of the three LODs, and labels in dynamic LODs as computed by our \emph{level-of-detail management}.
The dynamic LODs were chosen as the favorite approach by \rv{$40~\%$} of the participants. \rv{$53.33~\%$} of the participants preferred the \rv{highest} LOD. 
Participants, who selected the dynamic LODs as their favorite design, emphasized that the vertical stacking of labels is reduced while detailed information about close attractions is preserved. 
The participants who chose the \rv{highest LOD} as their favorite design appreciated the detailed information that can be used \rv{in} decision making. 
\rv{It is surprising} that they were not disturbed \rv{or annoyed} by the excessive vertical stacking of the labels.
The dynamic LODs avoid this excessive vertical stacking while presenting more information about close labels and less information about far labels. \rv{To check vertical stacking, Figure~\ref{fig:q5}(a)} \rv{compares} the \rv{highest} LOD and dynamic LODs. The more information is included for \rv{a label, the higher is the chance the label needs to be shifted upwards and stacked.}
We recorded the $y$-coordinate from the highest label of the two methods as a representative value for each themed area.
\rv{The height of the stacked labels} can be effectively reduced \rv{when} using the dynamic LODs.

For hypothesis \textbf{(H3)}, we \rv{aim to} compare the decision making effectiveness when using 2D \rv{paper maps} or our AR encoding in Task~3. 
We again \rv{measure} the task completion time and accuracy between using a Tokyo Disneyland map and our visualization. 
\rv{As a preprocessing, we first removed other POIs (e.g., shops or restaurants) and left the $35$ big attractions from the official 2D map of the amusement park, to increase the fairness of the comparison. More details about the task are included in the supplementary material.}
\rv{$60~\%$ of the participants selected attractions with minimal waiting times of a themed area when using the 2D map and $83.33~\%$ when the AR encoding was employed (Figure~\ref{fig:q5}(b)). The average time that the participants needed to select an attraction using the 2D map was $58.79~s$ while they spent $32.18~s$ on average when using our approach, which clearly shows a reduced effort for POI selection (Figure~\ref{fig:q5}(c))}.

\rv{In the feedback session, participants are allowed to freely comment on the \rv{presented} approach. Videos are shown highlighting the dynamic behavior of our tool when the user interacts with the system.}
Two participants mentioned that they prefer 2D maps compared to AR since 2D maps \rv{give a global top view. However, they performed the tasks in the user study better with the AR setting}. 
\rv{We believe that both 2D maps and AR systems} have strengths and weaknesses \rv{depending on the tasks and use cases. 
In our study, we have proven that for navigation purposes, AR systems could be more practical.} 
Two participants \rv{also suggested to combine 2D maps together with AR systems as done by Veas et al.~\cite{Veas:2012:TVCG}.
This could allow us to exploit the advantages of both approaches and achieve a similar result as in \emph{Google Maps} and \emph{Google Street View}.} 
% if our tool would be used for route planning. 
\rv{Other} participants would prefer super labels combined with the \rv{highest} LOD.
\rv{This could} reduce visual clutter\rv{, but might lead to a higher} vertical stacking of labels compared to dynamic LODs. \rv{We, therefore, allow users to adjust the thresholds for switching LODs, to accommodate this preference.} The occlusion handling and the smooth transitions \rv{were positively} mentioned by participants in the general feedback. 
Examples include: \emph{"I really like the occlusion management, to my eyes, it's almost seamless."} or, \emph{"Active occlusion handling is much superior to no occlusion handling."}. 
\rv{Super label aggregation has been another popular and specifically mentioned feature}. 
Participants appreciate the overview on the themed areas by giving feedback such as, \emph{"I like the super label transitions if there are many attractions because it gives a good overview of an area."}, \rv{and} \textit{"I like the super label transitions the most."}.  
\rv{Overall, all participants expressed interest to use our system for navigation purposes.}

%% file: sections/limitations.tex
\section{Limitations}
\label{ssec:limitations}

The limitations of our system are inherited from the \rv{hardware, especially the} accuracy of mobile GPS. 
The position and \rv{particularly} the rotation \rv{data from} the available Xiaomi Mi A2 smartphone and the Google Nexus C tablet \rv{are} not consistent \rv{based on our experience.} 
A less coherent behavior of our system \rv{follows} as the \rv{sensor data from} each of the two devices is not stable. \rv{This, unfortunately, limits the capability to fully utilize the application, while we also envision that} this will sooner or later be solved by \rv{newer technologies}. \rv{To remove the errors, we thus} present the results using predefined label positions \rv{in AR 3D world space.}  \rv{This allows us to avoid those errors induced by the hardware (e.g., changes in the device position and viewing angle) that could influence the coherence of labels. 
It will} be interesting to collaborate with researchers focusing on high-precision GPS positioning systems.

\rv{Another limitation is that the data could contain many POIs with long text descriptions. 
If each label should be large enough to show the text, not much background information could be \rv{depicted} eventually.
The current aggregation of labels to super labels is straightforward and can be easily extended based on the use cases.}
\rv{One important decision criterion for the \emph{occlusion management} and the \emph{level-of-detail management} is the position of the user. The ordering of the labels based on the position of the user influences the resulting labeling.}
\rv{Furthermore, one limitation is} the loss of the global overview using AR compared to 2D maps as \rv{mentioned by related work \cite{grassetimage, firstnavigation, guarese} and two user study participants}. Users need to interact with the system and look \rv{into} different directions to see all the labels. 
The AR view only depicts the labels that are currently in front of the user in the respective view volume. 
We could \rv{in the future} introduce additional labels on the sides of the screen to \rv{provide hints to invisible objects}.

%% file: sections/conclude.tex
\section{Conclusion and Future Work}
\label{sec:conclude}

We present a context-responsive labeling framework in Augmented Reality, which allows us to introduce rich-content labels associated with POIs.
The label management strategy suppresses label occlusions and incoherent label movements caused by transitions and rotations of the device during user interaction.
\rv{The framework} presents an alternative approach for spatial data navigation.
The \emph{level-of-detail management} takes the position of the user and label density in the view volume into account. The computed levels-of-detail for each label avoid excessive vertical stacking of labels, while still retaining basic information, which depends on the object distance. 
To further reduce visual clutter, we introduce the concept of super labels, which group a set of labels.
Smooth transitions have been implemented in our \emph{coherence management} to avoid flickering and enable stable label movement. The evaluation shows the applicability of the proposed approach.

As future direction, techniques will be investigated to overcome the drawbacks of seeing only the labels that are in the current view \rv{volume.} The user should still anticipate POIs outside the view volume and retain \rv{a} global overview of the annotated scene as with 2D maps. 
One possibility would be including the technique presented by Lin~et~al.~\cite{kaipaper} to depict labels that are currently outside the view volume \rv{and place hints at the display border} of the device. 
Considering the positioning accuracy, it would be interesting to include so-called \emph{Dual-Frequency GPS} \cite{dualfrequency} or \emph{Continuous Operating Reference Stations} (CORS) \cite{cors} as investigated by related work to improve the sensor accuracy of mobile devices \cite{kuhlmann}. %\cite{lowCost, kuhlmann, compass}. 
\rv{A selection scheme with the integration of service providers (e.g., OpenStreetMap or Google Maps with large POI data) could improve the system usability.}